\documentclass[10pt,journal,comsoc]{IEEEtran}
\usepackage{amsmath}
\usepackage{graphicx}
\usepackage{algorithm,float} 
\usepackage{algpseudocode}
\usepackage{caption,subcaption}
\usepackage{amsmath,amsfonts,amsthm,mdwmath} 
\usepackage{graphics,graphicx}   
\DeclareGraphicsExtensions{.eps,.pdf,.gif,.png,.jpg}
\usepackage{url}
\usepackage{algorithm}
\usepackage{multicol}
\usepackage{color}
\usepackage{booktabs}

\newcommand{\vect}[1]{\boldsymbol{#1}}

\begin{document}
\title{A Dispersed Federated Learning Framework for 6G-Enabled Autonomous Driving Cars}

\author{Latif~U.~Khan,~Yan~Kyaw~Tun,~Madyan~Alsenwi,~Muhammad~Imran,~\IEEEmembership{Member,~IEEE},~Zhu~Han,\IEEEmembership{~Fellow,~IEEE},~and~Choong~Seon~Hong,~\IEEEmembership{Senior~Member,~IEEE}

\IEEEcompsocitemizethanks{
\IEEEcompsocthanksitem L.~U.~Khan,Y.~.K.~Tun,~M.~Alsenwi,~and~C.~S.~Hong are with the Department of Computer Science \& Engineering, Kyung Hee University, Yongin-si 17104, South Korea.
\IEEEcompsocthanksitem M. Imran is with the College of Applied Computer Science, King Saud University, Riyadh, Saudi Arabia \& also with the Computer Science Department, University of Houston, Houston, TX 77004 USA, and the Department of Computer Science \& Engineering, Kyung Hee University, South Korea.
\IEEEcompsocthanksitem Zhu Han is with the Electrical and Computer Engineering Department, University of Houston, Houston, TX 77004 USA, and also with the Computer Science Department, University of Houston, Houston, TX 77004 USA, and the Department of Computer Science \& Engineering, Kyung Hee University, South Korea.

\IEEEcompsocthanksitem C. S. Hong is the correspoding author.

}
\thanks{
	}}

\markboth{}{}%

\IEEEcompsoctitleabstractindextext{%
\begin{abstract} 
Sixth-Generation ($6$G)-based Internet of Everything applications (e.g. autonomous driving cars) have witnessed a remarkable interest. Autonomous driving cars using federated learning (FL) has the ability to enable different smart services. Although FL implements distributed machine learning model training without the requirement to move the data of devices to a centralized server, it its own implementation challenges such as robustness, centralized server security, communication resources constraints, and privacy leakage due to the capability of a malicious aggregation server to infer sensitive information of end-devices. To address the aforementioned limitations, a dispersed federated learning (DFL) framework for autonomous driving cars is proposed to offer robust, communication resource-efficient, and privacy-aware learning. A mixed-integer non-linear (MINLP) optimization problem is formulated to jointly minimize the loss in federated learning model accuracy due to packet errors and transmission latency. Due to the NP-hard and non-convex nature of the formulated MINLP problem, we propose the Block Successive Upper-bound Minimization (BSUM) based solution. Furthermore, the performance comparison of the proposed scheme with three baseline schemes has been carried out. Extensive numerical results are provided to show the validity of the proposed BSUM-based scheme.

\end{abstract}

}
\maketitle
\IEEEdisplaynotcompsoctitleabstractindextext
\IEEEpeerreviewmaketitle

\section{Introduction}
\label{sec:introduction}
\setlength{\parindent}{0.7cm}The ongoing trend towards the development of Sixth-Generation ($6$G) wireless system will enable various emerging Internet of Everything (IoE) applications \cite{9163104,akyildiz20206g,yang20196g,khan2021digital}. These emerging IoE applications are brain-computer interaction, haptics, autonomous connected vehicles, among others \cite{saad2019vision}. Autonomous driving cars are expected to offer a wide variety of striking features in the foreseeable future because of the remarkable prevalence of emerging technologies. Autonomous driving cars can offer prominent smart safety features, e.g., traffic sign detection, lane departure warning, collision avoidance, and instant car accident reporting. Other than safe driving features, infotainment services using intelligent caching can be achieved within autonomous driving cars. Such features can be enabled via effective machine learning schemes \cite{grigorescu2019survey}. Several studies used centralized machine learning for enabling different services in autonomous driving cars \cite{ndikumana2020deep}. However, there is an inherent issue of privacy leakage in machine learning schemes based on centralized training. The reason for this is the transfer of devices data to the centralized server for training. Coping with the privacy leakage issue of centralized machine learning, \emph{federated learning (FL)} was introduced that is based on the training of a global machine learning model without transferring data, but only sends learning parameters from devices to a centralized server \cite{mcmahan2016communication}. Therefore, we can use FL in autonomous driving cars to offer various learning capabilities \cite{khan2019federated}. On the other hand, there are scenarios where continuous interaction with data-generating devices is necessary for training a machine learning model. Specifically, $4000$ gigaoctet of data is generated by cars every day \cite{FL_industry_1}. One time training of a centralized machine learning model might not produce good results. Additionally, autonomous cars generally have strict latency applications (e.g., autonomous cars accident reporting that require instant analytics at edge \cite{khan2019edge}). Therefore, edge intelligence (i.e., one of the key enabler of $6$G \cite{9163104}) and FL seems to be a promising solution for the training of a machine learning model for autonomous driving cars enabled by $6$G. \par
\setlength{\parindent}{0.7cm}Although FL offers various features, it has few serious concerns:\par
\begin{itemize}
\item \textit{Privacy leakage:} A malicious end-device/aggregation server can infer sensitive information of devices using the learning model updates \cite{khan2020federated, lim2019federated}. Therefore, we must ensure complete privacy preservation in FL. 

\item \textit{Robustness:} In a traditional FL, a global server aggregates the local learning models. FL process can be interrupted in case of a failure of the centralized aggregation server.

\item \textit{Centralized server security:} The traditional FL uses a frequent exchange of updates between the devices and an aggregation server. The centralized server can be attacked by a malicious entity that can alter the learning parameters, and disturbs the learning process.


\item \textit{Communication resources constraints:} The frequent exchange of updates between an aggregation server and the devices will require a significant communication resources which will impose limitations on the constrained communication resources. 
\end{itemize}\leavevmode\newline
To enable robust FL for autonomous driving cars over a wireless network, we can use blockchain-based FL \cite{kim2019blockchained}. In a typical blockchain-based FL, a devices set transmits the local model updates towards the miners installed at base stations (BSs). After trustful sharing of local models among miners, the blockchain is used for storing to learning model updates in a secure way. Finally, after completion of the blockchain consensus algorithm, every  BS transmits devices local models to its associated devices for global aggregation. However, reaching a consensus among miners have a high latency issue that is not desirable for autonomous driving car scenarios. Therefore, we can use distributed nodes (e.g., BS) for autonomous driving cars that can directly share learning model updates with each other via fast back-haul links without using blockchain and thus, avoid high-latency due to consensus algorithm. On the other hand, resource optimization in the FL process can be enabled via hierarchical learning fashion \cite{abad2019hierarchical}. However, the hierarchical federated learning has a robustness issue due to the use of centralized aggregation. Other than robustness and resource optimization, we must tackle the privacy leakage issue of FL. Differential privacy preservation scheme was proposed in \cite{seif2020wireless} to offer enhanced privacy preservation in FL by adding noise to local updates prior to their transmission. Differential privacy-based FL might suffer from prolonging in convergence time. Therefore, we must propose new schemes that must not prolong the FL convergence time.\par 
To cope with the aforementioned challenges, we propose a dispersed federated learning (DFL) framework for autonomous driving cars. The proposed framework offers enhanced robustness due to its decentralized aggregation nature. DFL is based on computing sub-global models within various groups (e.g., people within autonomous cars). The users of a group perform learning of their local models and send them to access points within cars for sub-global aggregations. All the end-devices downloads the sub-global models and update their local models. This iterative process is carried out for fixed sub-global iterations. The next step is share the sub-global models among various cars using roadside units (RSUs). Finally, global aggregation takes place at all RSUs which sends the model to the cars. Furthermore, our DFL offers us with the additional advantage of better privacy preservation compared to the traditional FL. One can easily infer sensitive information of end-devices from local models (i.e., at the sub-global servers). On the other hand, it is extremely difficult for a global server to infer devices information from sub-global model updates  \cite{khan2020dispersed}. Other than privacy, the reuse of already occupied resource blocks by cellular users is more feasible with low transmit power due to close vicinity nature of the devices used for sub-global model computation. To the best of our knowledge, we are first to consider DFL in autonomous driving cars to offer robust, privacy-aware, and resource optimization. \par
\setlength{\parindent}{0.7cm}Our contributions are summarized as follows:\par
\begin{itemize}
    \item We present a novel DFL framework for autonomous driving cars that uses distributed learning fashion in the training of the FL model to offer robustness, efficient communication resources utilization, and enhanced privacy preservation. 
   
    \item A mixed-integer non-linear (MINLP) problem is formulated to minimize the global learning cost for the proposed DFL scheme. The formulated cost considers latency and loss in global DFL model accuracy due to packet error rate (PER). Furthermore, flexibility of tuning between error rate and latency is provided by the cost function. Our formulated problem minimizes DFL cost by optimizing resource allocation, power allocation, and association of autonomous cars to RSUs. Our problem remains non-convex even after relaxing resource allocation and association variables from binary to continuous variables. Therefore, we cannot use traditional convex optimization schemes.We propose Block Successive Upper-bound Minimization (BSUM) based solution. BSUM is a power optimization tool for the non-convex, non-smooth problems.

    \item Finally, numerical results are presented to show the effectiveness of the proposal.       
    
\end{itemize}\par
The rest of the paper is organized as follows. Section~\ref{Related Works} discusses related works regarding resource optimization in FL, BSUM-based schemes for resource allocation, association, and power allocation. Section~\ref{DFL} presents the proposed dispersed federated learning framework. System model and problem formulation are given in Section~\ref{sec:systemmodel}. The proposed solution is given in Section~\ref{BSUM solution}. Section~\ref{numerical results} presents numerical results. Finally, the paper is concluded in Section~\ref{conclusion}.

\section{Related Works} 
\label{Related Works}
\setlength{\parindent}{0.7cm}In this paper, we focus on resource optimization in wireless FL. In \cite{khan2019federated}, Khan \textit{et al.} surveyed incentive mechanism and resource optimization in FL for edge networks. In another study \cite{dinh2019federated}, Tran \textit{et al.} discussed the optimization model and analysis of wireless FL. The authors considered a problem to jointly minimize computing time and energy consumption for FL. Additionally, two trade-offs such as (a) energy consumption of device and FL model computation time, and (b) computation and latencies in communication as per the accuracy level of learning, were discussed. In \cite{chen2019joint}, Chen \textit{et al.} studied FL over wireless networks. Specifically, loss in global FL model accuracy due to channel uncertainties were studied. A problem was considered to optimize transmit power, user selection, and resource allocation for reducing the packet error error in wireless FL. The authors provided a detailed analysis of PER on the wireless FL accuracy and proved the closed-form expression of the expected FL convergence rate. Using the derived closed-form expression, the optimization problem is simplified and is solved via the Hungarian algorithm. Furthermore, the authors discussed the implementation complexity of their proposed scheme. In another study \cite{wang2019adaptive}, Wang \textit{et al.} considered FL for edge networks. A control algorithm under resource budget constraints that offers a trade-off between global parameter aggregation and local update for minimizing the loss function, was proposed. Validation of their algorithm was carried out using a real dataset through simulation results and prototype implementation. \par 
On the other hand, a hierarchical FL was proposed in \cite{abad2019hierarchical}. Their system model was based on a single MBS and several SBS. Initially, the sub-global model is computed at every SBS, which is then sent to the MBS. To further enhance the communication efficiency, the authors considered gradient sparsification and periodic averaging in their proposed hierarchical FL model. The authors validated their model using CIFAR-$10$ dataset which showed a significant increase in communication efficiency with almost similar learning model accuracy as compared to traditional FL. All the works considered in \cite{dinh2019federated,chen2019joint,wang2019adaptive, abad2019hierarchical} is based on a single server for performing aggregation that might fails due to a security attack or physical damage. Additionally, they did not tackle the privacy leakage issue in FL. In contrast to the work presented in \cite{dinh2019federated,chen2019joint,wang2019adaptive}, our proposed scheme offers robustness operation by truly adopting decentralization (will be explained in more detail in Section~\ref{DFL}). Furthermore, DFL reuses the resource blocks other users, and thus offers efficient resource management. \par 
\setlength{\parindent}{0.7cm}In \cite{kim2019blockchained}, Kim \textit{et al.} presented an idea of blockchain-based FL. First, local model computation takes place at a set of devices and sending to the miners. The job of the miners is to exchange local model updates in a secure way using blockchain and cross verification. Next, a block is generated having local model updates and send back to devices by the miners. Finally, a global update takes place on every device. The authors performed an end-to-end latency analysis and validated their proposal using numerical results. Although blockchain-based FL can offer trustful verification of wireless FL devices, the blockchain consensus algorithm has an inherent high latency.  \par 

Many works considered BSUM for various purposes in wireless networks \cite{ndikumana2019joint,tun2020joint,hong2015unified}. In \cite{hong2015unified}, BSUM was introduced that is based on a generalization of block coordinate descent (BCD). BSUM expands the application of BCD to a broader prospect. For BCD, it's difficult to solve non-convex optimization problems. On the other hand, BSUM can be applied easily for non-convex problems. In \cite{tun2020joint}, tun \textit{et al.} considered a joint content caching and resource allocation for virtualized wireless networks. They have formulated a problem to minimize the overall delay experienced by mobile virtual network operators end-users. They used BSUM for resource allocation, power control, and cache decision while minimizing the overall delay experienced by MVNO end-users. Another work \cite{ndikumana2019joint} considered control, caching, computation, and communication in mobile edge computing. To jointly optimize network latency and bandwidth usage, the authors formulated an optimization problem. Due to the non-convex nature of the formulated problem, the authors proposed a BSUM-based solution was proposed to minimize the latency and bandwidth usage by optimizing computational offloading (i.e., from end-devices to edge server and between edge servers) and data caching. Their proposed BSUM-based schemes showed promising results. In this paper, we use a BSUM-based solution to solve our formulated optimization because of its effectiveness in solving complex optimization non-convex optimization problems \cite{ndikumana2019joint,tun2020joint,hong2015unified}.\par

\begin{figure*}[!t]
	\centering
	\captionsetup{justification=centering}
	\includegraphics[width=18cm, height=10cm]{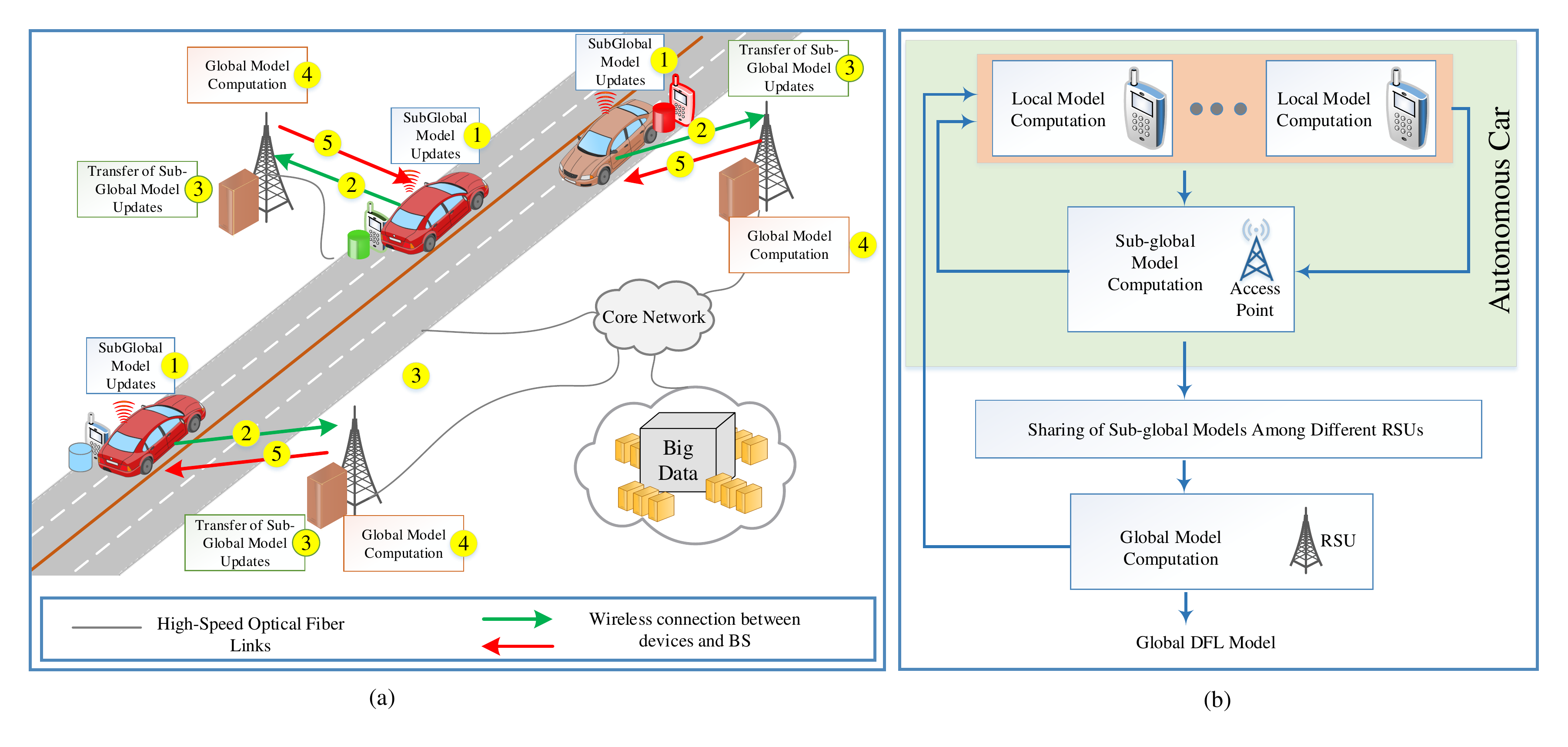}
	\caption{Dispersed federated learning enabled autonomous driving cars: (a) System overview, (b) Sequence diagram}
	\label{fig:system_model}
\end{figure*}

\section{Dispersed Federated Learning for Autonomous Cars}
\label{DFL}
In this section, we present our DFL framework for vehicular networks. As mentioned earlier that autonomous cars generate a significant amount of data every day. Therefore, using centralized machine learning with one-time training might cause performance degradation. Additionally, sending an enormous amount of generated data will consume significant communication resource. Therefore, FL can be a promising candidate for autonomous driving cars. However, FL has its own challenges. These challenges are robustness, wireless security, high communication resource consumption for training. To enable FL with robustness, wireless security, and communication resource-efficient, we propose a DFL framework for autonomous driving cars. The proposed DFL (presented in Fig.~\ref{fig:system_model}b) has the following steps:\par

\begin{itemize}
    \item \textbf{Step 1: Sub-global model computation:} First of all, every device located within autonomous driving cars computes its local model. Within autonomous car, the local models computed by devices are sent to their corresponding access points, where aggregation of the local model takes place for obtaining a sub-global model. The sub-global model is then sent back to the associated devices and thus, the sub-global model is updated in an iterative manner. In our paper, we assume a fixed iterations for computing a sub-global model for all autonomous driving cars. In the proposed DFL framework, we can have learning model weights deviation at the sub-global level and global level. These learning model weights deviation is strictly dependent on the data heterogeneity fashion \cite{Client-Edge}. To tackle this issue of weight deviations due to data heterogeneity, one must use an effective aggregation scheme. 

    \item \textbf{Step 2: Autonomous driving cars and roadside units association:} The autonomous driving cars are associated with the RSUs based on some effective criteria (will be explained later in the paper). It must be noted here that we can easily reuse wireless channel resources occupied by cellular users for communication between autonomous cars and RSUs. Therefore, we can say that our proposed DFL will offer us with efficient reuse of wireless channel resources.  
    \item \textbf{Step 3: Sub-global model updates transfer:} After the RSU receives the sub-global model updates from the autonomous driving cars, they transfer them to the other RSUs via a high-speed back-haul link. Here, we can notice that RSU cannot infer sensitive information of devices from sub-global updates. This is one of advantages of DFL over traditional FL.  
    \item \textbf{Step 4: Global FL model computation:} After the global model computation, every RSU performs global DFL model aggregation. The advantage of performing aggregation of sub-global models at different RSUs is robustness in contrast to traditional FL that is based on centralized aggregation. \par
    \item \textbf{Step 5: Dissemination of global model updates:} Finally, RSUs send the global models to cars, which disseminates them to their devices.\par
\end{itemize}\par
\begin{algorithm}[t!]
	
	\caption{Dispersed Federated Learning Algorithm}
	\label{algo:dispersedfederatedlearning}
	
	\begin{algorithmic}[1]
		\State Weights initialization $\boldsymbol{\omega}_{0}$
		\For {i=1,2,..., Global Rounds}
		\State \textbf{\emph{Step 1: Parallel run for all cars}}
        \For {k = 1,2,..., Sub-global iterations}
            \For{i=1,2,..., Local iterations}
                \State $\boldsymbol{\omega}_{n}(i) \longleftarrow \boldsymbol{\omega}_{n}(i-1)- \eta \nabla F_{n}(\boldsymbol{\omega}_{n}(i-1))$
            \EndFor
            \State $\boldsymbol{\omega}_{s}(k)\longleftarrow~Sub-global~aggregation({\boldsymbol{\omega}_{n}})$
            \State $\boldsymbol{\omega}_{n} \longleftarrow \boldsymbol{\omega}_{s}$
        \EndFor
        \State \textbf{\emph{Step 2: Global aggregation at RSUs}}
         \State $\boldsymbol{\omega}_{g}(k)\longleftarrow~global~Aggregation({\boldsymbol{\omega}_{s}})$
            \State $\boldsymbol{\omega}_{n} \longleftarrow \boldsymbol{\omega}_{g}$
        \EndFor 
	\end{algorithmic}
\end{algorithm} \par

The summary of the proposed DFL algorithm for autonomous cars is given in \textit{Alg.~\ref{algo:dispersedfederatedlearning}}. First of all, weights are initialized at access points of autonomous cars which share them with their associated devices (line-1). Next, all the cars compute their sub-global models in a parallel fashion (lines 2-10). Within every car, local learning models are computed and shared with access points to yield a sub-global model. This sharing of learning models between end-devices and access points is in an iterative manner (lines 4-7). The sub-global models are sent to the RSUs next to sub-global model computation. Finally, RSUs share learning model updates with each other and perform global model computation at every RSU (line-12). \par

\section{System Model and Problem Formulation}
\label{sec:systemmodel}
\setlength{\parindent}{0.7cm}Consider a network that consists of a set $\mathcal{M}$ of $M$ edge computing server-based RSUs and a set $\mathcal{N}$ of $N$ autonomous driving cars. Both autonomous driving cars and RSUs are located in an area covered by a base station (BS). The set of $Y$ cellular users served by BS is denoted by $\mathcal{Y}$. To offer effective frequency reuse-ability, the same set $\mathcal{R}$ of $R$ orthogonal resource blocks are used by both cellular users and autonomous driving cars for communication with the BS and RSUs, respectively. Within every autonomous driving car, a set $\mathcal{U}_{n}$ of $U_{n}$ users with local datasets want to train a global FL model.  
\setlength{\parindent}{0.7cm}Next, we discuss FL over wireless networks (Sub-section~\ref{FEDLm}), communication model (Sub-section~\ref{communication_model}), and problem formulation (Sub-section~\ref{Problem Formulation}) in the following subsections, respectively. \par
\subsection{Federated Learning Model} 
\label{FEDLm}
\setlength{\parindent}{0.7cm} A typical FL process has a devices set that performs learning of their local models. These local models are transmitted to an aggregation server for global aggregation. Finally, the devices download the global model to update their local model. In our model, there are a number of autonomous driving cars which communicate with the RSUs for performing learning. Every device $u_{n}$ of autonomous car $n$ has a local dataset $\mathcal{D}_{u}^{n}=[\boldsymbol{d}_{u_{1}}^{n}, \boldsymbol{d}_{u_{2}}^{n}, ..., \boldsymbol{d}_{u_{k_{u}^{n}}}^{n}]$, where $k_{u}^{n}$ represents the total number of data samples in the local dataset of device $u_{n}$ of car $n$. The type of the FL task determines the size of input sample $\boldsymbol{d}_{u_{k}}^{n}$ and its output $ \Theta_{u_{k}}^{n}$. The output $ \Theta_{u_{k}}^{n}$ is determined by weights $\boldsymbol{w}_{u}^{n}$ for a given input $\boldsymbol{d}_{u_{k}}^{n}$. In our model, we assumed a single output for a particular input sample. All the users are assumed to have different dataset sizes and distribution to reflect practical scenarios \cite{khan2019federated}. For instance, for a linear regression the output is given by $\Theta_{u_{k}}^{n}=\boldsymbol{w}_{u}^{n}\boldsymbol{d}_{u_{k}}^{n}$. FL is aimed to minimize global loss function $f$, i.e.,\par
    \begin{subequations}\label{eq:formluationF}
	\begin{align}
	 &\underset{\vect{\boldsymbol{w}_{1}^{n},\boldsymbol{w}_{2}^{n},..., \boldsymbol{w}_{U_{n}}^{n}}}{\text{min}} \frac{1}{K} \sum_{n=1}^{N}\sum_{u=1}^{U_{n}}\sum_{k=1}^{k_{u}^{n}}{f(\boldsymbol{w}_{u}^{n},\boldsymbol{d}_{u_{k}}^{n},\Theta_{u_{k}}^{n})},\label{federated_amin}\\
		     & s.t. \boldsymbol{w}_{u}^{1}=\boldsymbol{w}_{u}^{2}=...=\boldsymbol{w}_{u}^{N}=\boldsymbol{z}, \forall{u \in \mathcal{U}_{n}},\forall{n \in \mathcal{N}},\label{federated_bmin}
	\end{align}
	\end{subequations}\\
where $\boldsymbol{z}$ and $K$ are the global model and size of training data for all devices of all autonomous driving cars, respectively. The loss function $f$ depends on the application nature. For instance, for prediction it counts for prediction error. For linear regression problem, the loss function is given by $f(\boldsymbol{w}_{u}^{n},\boldsymbol{d}_{u_{k}}^{n},\Theta_{u_{k}}^{n})=\frac{1}{2}(\boldsymbol{d}_{u_{k}}^{n}\boldsymbol{w}_{u}^{n}-\Theta_{u_{k}}^{n})^{2}, \forall{n\in \mathcal{N}}$. The constraint (\ref{federated_bmin}) restricts that all the devices must have the same FL model. On the other hand, the global model is given by:\par
           \begin{equation}
            \label{eq:1}
		    \begin{aligned}
		    \boldsymbol{z} = \frac{\sum_{n=1}^{N}\sum_{u=1}^{U_{n}}{k_{u}^{n}\boldsymbol{w}_{u}^{n}}}{K}.
		    \end{aligned}
		    \end{equation}\par
\setlength{\parindent}{0.7cm}Mainly, there are two kinds of resources, i.e., computational and communication resources, required for training of the global FL. It is necessary to optimally use these resources for training. Furthermore, wireless FL has a performance degradation due to uncertainties in the wireless channel \cite{chen2019joint}. The PER for device $u_{n}$ of autonomous car $n$ over resource block $x_{n,r}$ and waterfall threshold $\vartheta$ is given by \cite{xi2011general}:\par
            \begin{equation}
            \label{eq:2}
		    \begin{aligned}
		    q_{u,n}(\boldsymbol{A},\boldsymbol{X},\boldsymbol{P}) =a_{n, m}x_{n,r}\left( 1-\exp{\left(\frac{-\vartheta (\sum_{y \in \mathcal{Y}_{r}}h_{y}^{r}P_{y}^{r}+\sigma^{2})}{p_{n}h_{n,m}^{r}}\right)}\right),
		    \end{aligned}
		    \end{equation}\newline
where $x_{n,r}$ and $a_{n, m}$ denote the resource block association and autonomous car-RSU association variables, respectively. $p_{n}$ denotes the transmit power of device $n$. The transmission power of the autonomous cars should be selected within the limit.\par
        \begin{equation}
        \label{eq:3}
         0 \leq p_{n} \leq P_{m}, \forall{n \in \mathcal{N}} 
        \end{equation}\par
On the other hand, the total transmission power of all autonomous driving cars must be less than or equal to the maximum allowed transmission power threshold.
        \begin{equation}
        \label{eq:4}
         \sum_{n=1}^{N}{p_{n}} \leq P_\textrm{max}^{\textrm{trans}}. 
        \end{equation}\par

To assign resource blocks to autonomous driving cars, we use binary variable $x_{n,r}$:\par
        \begin{equation}
        \label{eq:5}
        x_{n,r} = \begin{cases}
        1, & \text{If car $n$ is assigned resource block $r$},\\
        0, & \text{otherwise}. 
        \end{cases}
        \end{equation}\par
\setlength{\parindent}{0.7cm}The association of autonomous car $n$ with RSU $m$ is given by the binary variable $a_{n, m}$:\par
        \begin{equation}
        \label{eq:6}
        a_{n, m} = \begin{cases}
        1, & \text{If car $n$ is associated with RSU $m$},\\
        0, & \text{otherwise}. 
        \end{cases}
        \end{equation}\par
\setlength{\parindent}{0.7cm}The number of autonomous driving cars associated with a RSU $m$ is restricted by a maximum number $\Delta_{m}$, i.e.,\par
            \begin{equation}
            \label{eq:7}
		    \begin{aligned}
		    \sum_{n=1}^{N} {a_{n, m}}  \leq \Delta_{m}, \forall{m \in \mathcal{M}}, 
		    \end{aligned}
		    \end{equation}\newline
where $\Delta_{m}$ denotes the maximum autonomous driving cars that can be served by RSU $m$ simultaneously. To study the effect of PER on FL, we first consider the case of traditional FL. We use the binary variable $\Pi_{u}^{n}$ to indicate whether the received local model from device $u_n$ of the autonomous car $n$ contains errors or not ($\Pi_{u}^{n}=1$ if it does not contain errors and $0$ otherwise). To capture the effect of PER on traditional FL model, (\ref{eq:1}) can be re-written as:\par
            \begin{equation}
            \label{eq:8}
		    \begin{aligned}
		    \boldsymbol{z}= \frac{\sum_{n=1}^{N}\sum_{u=1}^{U_{n}}{k_{u}^{n}\boldsymbol{w}_{u}^{n}}\Pi_{u}^{n}}{\sum_{n=1}^{N}\sum_{u=1}^{U_{n}}{k_{u}^{n}}\Pi_{u}^{n}}.
		    \end{aligned}
		    \end{equation}\par
\setlength{\parindent}{0.7cm}Let $F(\boldsymbol{z})=\frac{1}{K}\sum_{n=1}^{N}\sum_{u=1}^{U_{n}}\sum_{k=1}^{k_{n}^{u}}{f(\boldsymbol{z},\boldsymbol{d}_{uk}^{n},\Theta_{uk}^{n})}$. Mainly, PER degrades the wireless FL performance. Therefore, to compute expression for the cost function that counts for the effect of PER on FL, we made several assumptions \cite{chen2019joint, dinh2019federated}:\par      

\begin{itemize}
    \item It is assumed that the gradient $\bigtriangledown(F(\boldsymbol{z}))$ with respect to $\boldsymbol{z}$ is uniform Lipschitz continuous \cite{friedlander2012hybrid}. 
           \begin{equation}
            \label{eq:9}
		    \begin{aligned}
		    \parallel \bigtriangledown(F(\boldsymbol{z}_{t+1})) -\bigtriangledown(F(\boldsymbol{z}_{t}))    \parallel   \leq L \parallel \boldsymbol{z}_{t+1}- \boldsymbol{z}_{t} \parallel
		    \end{aligned}
		    \end{equation}\par
		    where $L$ is positive constant, whereas $\parallel*\parallel $ represent norm of $*$.   
    \item For $\mu > 0$, $F(\boldsymbol{z})$ is strongly convex:
            \begin{equation}
            \label{eq:10}
		    \begin{aligned}
		    F(\boldsymbol{z}_{t+1})\geq  F(\boldsymbol{z}_{t})+(\boldsymbol{z}_{t+1}-\boldsymbol{z}_{t})^{T}\bigtriangledown F(\boldsymbol{z}_{t})+\frac{\mu}{2} \parallel\boldsymbol{z}_{t+1}-\boldsymbol{z}_{t}\parallel^{2}
		    \end{aligned}
		    \end{equation}\par
    \item $F(\boldsymbol{z})$ is  twice-continuously differentiable function. Using $(\ref{eq:9})$ and $(\ref{eq:10})$ we can write:
           \begin{equation}
            \label{eq:11}
		    \begin{aligned}
		    \mu \boldsymbol{I} \preceq \bigtriangledown^{2}F(\boldsymbol{z}) \preceq L\boldsymbol{I}
		    \end{aligned}
		    \end{equation}\par
    \item For positive constants $\zeta_{1}\geq 0$ and $
    \zeta_{2}\geq 0$, we have $\parallel \bigtriangledown f(\boldsymbol{z}_{t},d_{uk}^{i},\Theta_{uk}^{i})\parallel^{2} \leq  \zeta_{1}+\zeta_{2}  \parallel \bigtriangledown F(\boldsymbol{z}_{t}) \parallel^{2} $. \par
    \end{itemize}\par
\setlength{\parindent}{0.7cm}For a FL loss functions (i.e., logistic or linear loss functions) fulfilling the above assumptions and exactly one resource block per autonomous car, then the effect of PER on FL model is given by (\ref{eq:2}). Therefore, we consider the impact of PER on global model computation for DFL using the cost function $C_{p}$ as \cite{chen2019joint}.
           \begin{equation}
            \label{eq:12a}
		    \begin{aligned}
		    C_{p}(\boldsymbol{A},\boldsymbol{X},\boldsymbol{P})=\sum_{n=1}^{N}\ q_{n}(\boldsymbol{A},\boldsymbol{X},\boldsymbol{P}),
		    \end{aligned}
		    \end{equation}\\
where $q_{n}=a_{n, m}x_{n,r}\left( 1-\exp{\left(\frac{-\vartheta (\sum_{y \in \mathcal{Y}_{r}}h_{y}^{r}P_{y}^{r}+\sigma^{2})}{p_{n}h_{n,m}^{r}}\right)}\right)$. Next, we discuss the channel model.

\subsection{Communication Model}
\label{communication_model}
In our system model, orthogonal frequency division multiple access (OFDMA) is considered. Different resource blocks are associated with different autonomous driving cars, and thus there will be no interference between autonomous driving cars. However, we assume that the set orthogonal resource blocks $\mathcal{R}$ are reused, and thus there will be interference between devices and cellular users. In our work, one resource block can be allocated to a single car:\par
           \begin{equation}
            \label{eq:13}
		    \begin{aligned}
		    \sum_{n=1}^{N} {x_{n,r}}  \leq 1, \forall{r \in \mathcal{R}}.
		    \end{aligned}
		    \end{equation}\par
\setlength{\parindent}{0.7cm}Every autonomous car must not get more than one resource block:\par            
            \begin{equation}
            \label{eq:14}
		    \begin{aligned}
		    \sum_{r=1}^{R} {x_{n,r}}  \leq 1, \forall{n \in \mathcal{N}}. 
		    \end{aligned}
		    \end{equation}\par
\setlength{\parindent}{0.7cm}The upper limit on resource blocks assigned to the autonomous driving cars can be given by.\par
            \begin{equation}
            \label{eq:15}
		    \begin{aligned}
		    \sum_{n=1}^{N} \sum_{r=1}^{R}{x_{n,r}}  \leq R.
		    \end{aligned}
		    \end{equation}\par
\setlength{\parindent}{0.7cm}Let the up-link channel gain between the autonomous driving cars $n\in \mathcal{N}$ and RSU $m$ for $r$ can be given by $h_{n,m}^{r}$ and assumed constant during the transmission of learning parameters. The signal-to-interference-plus-noise ratio (SINR) of autonomous car $n$ is given by:\par
            \begin{equation}
            \label{eq:16}
		    \begin{aligned}
		    \Gamma_{n} = \frac{p_{n}h_{n,m}^{r}}{\sum_{y \in \mathcal{Y}_{r}}h_{y}^{r}P_{y}^{r}+\sigma^{2}},
		    \end{aligned}
		    \end{equation}\newline
where $ p_{n} $ and $\sigma^{2}$ denote the power transmitted by autonomous car $n$ associated with RSU $m$ and noise, respectively. The term $\sum_{y \in \mathcal{Y}_{i}}h_{y}^{r}P_{y}^{r}$ denote the interference for resource block $r$ on autonomous car because of cellular users. The variables $h_{y}^{r}$ and $P_{y}^{r}$ denote the transmitted power of the $y$ cellular user and channel gain between the BS and $y$ cellular user for a resource block $r$, respectively. Furthermore, we assume that all autonomous vehicles transmit signals with equal power. The up-link achievable data rate for the autonomous car $n$ is given by: \par
            \begin{equation}
            \label{eq:17}
		    \begin{aligned}
		    \eta_{n} = \Omega_{n,m}^{r}\log_{2}(1+\Gamma_{n}),
		    \end{aligned}
		    \end{equation}\newline
where $\Omega_{n,m}^{r}$ is the allocated bandwidth to the autonomous car $n$ associated with RSU $m$ for a resource block $r$. We assume that all the resource blocks have equal bandwidth. The transmission delay occurred in sending the sub-global model updates of the autonomous car $n$ to RSU $m$ is given by:\par
            \begin{equation}
            \label{eq:18}
		    \begin{aligned}
		    T_{n}^{\textrm{trans}}(\boldsymbol{A},\boldsymbol{X},\boldsymbol{P}) = \frac{Q_{n}x_{n,r}a_{n, m}}{\eta_{n}}. 
		    \end{aligned}
		    \end{equation}\newline
where $Q_{n}$ denote the size of sub-global model updates of the autonomous car $n$.

\subsection {Problem Formulation}
\label{Problem Formulation}
\setlength{\parindent}{0.7cm} This sub-section discusses problem formulation whose purpose is to jointly minimize computation time and PER. From (\ref{eq:12a}), it is clear that the PER has a proportional effect on DFL model error. Therefore, we get the intuition of $C_{p}$ to reflect PER on a global DFL model in our system model. The total cost for DFL is given by: \par
		    \begin{multline}
		    \label{eq:19}
		    C_\textrm{Global}(\boldsymbol{A},\boldsymbol{X},\boldsymbol{P})=\alpha (\sum_{n=1}^{N}q_{n}(\boldsymbol{A},\boldsymbol{X},\boldsymbol{P}))+  \beta (\sum_{n=1}^{N}T_{n}^{\textrm{trans}}(\boldsymbol{A},\boldsymbol{X},\boldsymbol{P})),\\
		    \end{multline}\newline
where $\alpha$ and $\beta$ are the constants and their sum is $\alpha+\beta=1$. We formulate our problem joint resource allocation, transmission power allocation, and autonomous driving cars association problem (\textbf{P1}) to minimize the cost associated with DFL as follows: \par
    \begin{subequations}\label{eq:problem_formulation}
	\begin{align}
   \mathbf{P1}:\ \	&\underset{\vect{\textbf{A}}, \vect{\textbf{X}},\vect{\textbf{P}}}{\text{min}}\ \  C_\textrm{Global}(\boldsymbol{A},\boldsymbol{X},\boldsymbol{P})
	\tag{\ref{eq:problem_formulation}}\\
	&\text{subject to:}\nonumber\\
	& 0 \leq p_{n} \leq P_{m}, \forall{n \in \mathcal{N}},
	\label{first:a}\\
	&  \sum_{n=1}^{N}{p_{n}} \leq P_\textrm{max}^{\textrm{trans}} ,
	\label{first:b}\\
	& \sum_{n=1}^{N} {x_{n,r}}  \leq 1, \forall{r \in \mathcal{R}} ,
	\label{first:c}\\
	&   \sum_{m=1}^{M} {a_{n, m}}  \leq 1, \forall{n \in \mathcal{N}} ,
	\label{first:d}\\
	&\sum_{r=1}^{R} {x_{n,r}}  \leq 1, \forall{n \in \mathcal{N}},
	\label{first:e}\\
    &\sum_{n=1}^{N} \sum_{r=1}^{R}{x_{n,r}}  \leq R,\label{first:f}\\
    & \sum_{n=1}^{N} {a_{n, m}}  \leq \Delta_{m}, \forall{m \in \mathcal{M}}, \label{first:g}\\
	&a_{n, m}\in \{0,1\}~~~~\forall{n \in \mathcal{N}, m \in \mathcal{M}},\label{first:h}\\
	& x_{n,r}\in \{0,1\}~~~~\forall{n \in \mathcal{N}, m \in \mathcal{M}}.\label{first:i}
	\end{align}
    \end{subequations}\par

\setlength{\parindent}{0.7cm}Problem \textbf{P1} is to minimize the total cost of one global DFL model computation. Constraint (\ref{first:a}) restricts the transmission power assignment. The total transmission power for all devices should not exceed the upper limit (\ref{first:b}). Constraint (\ref{first:c}) restricts resource blocks allocation. Constraint (\ref{first:d}) shows that the association of an autonomous car can be made to a maximum of one RSU. Every autonomous driving car must be not getting more than one resource block according to the constraint (\ref{first:e}). Constraint (\ref{first:f}) shows that the assigned resource blocks to cars must fulfill the upper limit. The maximum number of autonomous driving cars that can be associated to a particular RSU is restricted by the constraint (\ref{first:g}). Constraints (\ref{first:h}) and (\ref{first:i}) restricts that association variable $a_{n, m}$ and resource block assignment variable $x_{n,r}$  to binary values. Problem \textbf{P1} has combinatorial nature and it becomes NP-hard for large devices and resource blocks. Problem \textbf{P1} remains non-convex even if we transform the binary variables $\textbf{A}$ and $\textbf{X}$ into continuous variables, therefore, we cannot use BCD algorithm. To cope with the aforementioned challenges associated with problem \textbf{P1}, we use the BSUM-based method. \par

\section{BSUM-Based DFL Cost Minimization}
\label{BSUM solution}
Solving Problem \textbf{P1} using convex optimization schemes is very difficult due to its combinatorial nature. Therefore, we use BSUM to solve Problem \textbf{P1} \cite{tun2020energy}. Next, we present an overview of BSUM (Sub-section~\ref{bsumoverview}) and BSUM-based algorithm (Sub-section~\ref{proposed algorithm}) to solve Problem \textbf{P1}.  

\subsection{BSUM Overview}
\label{bsumoverview}
BSUM is based on parallel computing in a distributed manner. In contrast to centralized schemes, BSUM offers efficient problem decomposability and faster convergence \cite{han2017signal}. The standard form of BSUM can be given by.

   \begin{equation}
    \label{eq:20b}
	\begin{aligned}
	 \underset{\vect{\boldsymbol{v}}}{\text{min}} ~g(\boldsymbol{v}_{1},\boldsymbol{v}_{2},...,\boldsymbol{v}_{J}), s.t. \boldsymbol{v}_{j} \in \mathcal{Z}_{j}, \forall{j \in \mathcal{J}, j=1,2,..,J},
	\end{aligned}
	   \end{equation}\newline
where $\mathcal{Z}:=\mathcal{Z}_{1} \times \mathcal{Z}_{2}\times ..., \mathcal{Z}_{J}$ and $g(.)$ is a continuous function. $\boldsymbol{v}_{j}$ and $\mathcal{Z}_{j}$ are a block of variables and closed convex set, respectively. A single block of variables can be solved using the BCD algorithm by optimizing the following problem. 
    \begin{equation}
    \label{eq:21ba}
	\begin{aligned}
	 \boldsymbol{v}_{j}^t \in \underset{\vect{\boldsymbol{v}_{j}\in \mathcal{Z}_{j}}}{\text{argmin}} ~g(\boldsymbol{v}_{j},\boldsymbol{v_{-j}^{t-1}}),
	\end{aligned}
	\end{equation}\newline
where $\boldsymbol{v}_{-j}^{t-1}:= (v_{1}^{t-1},..., v_{j-1}^{t-1}, v_{j+1}^{t-1},..., v_{j}^{t-1}), \boldsymbol{v}_{k}^{t}=\boldsymbol{v}_{k}^{t-1}$ for $j\neq k$. It is difficult to solve both (22) and (23). If (22) has non-convex nature then BCD might not always converge. To overcome this issue, one can use BSUM that is based on using proximal upper-bound function $h(\boldsymbol{v}_{j}, \boldsymbol{y})$ of $g(\boldsymbol{v}_{j}, \boldsymbol{y}_{-j})$. Jensen's upper-bound, linear upper-bound, and quadratic upper-bound are various upper-bound that can be used in BSUM \cite{han2017signal}. Upper-bound function must satisfy the following assumptions.

\textbf{Assumptions:} \textit{The following assumptions are used in our work.}\par
\textit{1) $h(\boldsymbol{v}_{j},\boldsymbol{y})= g(\boldsymbol{y})$} \par
\textit{2) $h(\boldsymbol{v}_{j},\boldsymbol{y}) > g(\boldsymbol{v}_{j}, \boldsymbol{y}_{-j})$, }\par
\textit{3) $h'(\boldsymbol{v}_{j},\boldsymbol{y}; \boldsymbol{q}_{j})|_{\boldsymbol{v}_{j}=\boldsymbol{y}_{j}}= g'(\boldsymbol{y};\boldsymbol{q}), \boldsymbol{y}_{j}+\boldsymbol{q}_{j} \in \mathcal{Z}_{j}$,}\\
where $h$ denotes the proximal upper-bound function. The global upper-bound nature of $h$ is guaranteed by assumptions $(1)$ and $(2)$. Additionally, the steps of the upper-bound function $h(\boldsymbol{v}_{i},\boldsymbol{y})$ are negative of objective function gradient in the direction of $q$. In this paper, we use an upper-bound that is applied to the objective function via quadratic penalization.
    \begin{equation}
    \label{eq:22}
	\begin{aligned}
	h(\boldsymbol{v}_{j},\boldsymbol{y})=g(\boldsymbol{v}_{j},\boldsymbol{y}_{-j})+\frac{\mu}{2} \parallel \boldsymbol{v}_{j}-\boldsymbol{y}_{j} \parallel^{2},
	\end{aligned}
	\end{equation}\\
where $\mu$ is positive penalty constant. The proximal upper-bound is solved using the following update rule.

\begin{equation}
 \label{eq:23}
    \begin{cases}
      \boldsymbol{v}_{j}^{t} \in \underset{\vect{\boldsymbol{v}_{j}\in \mathcal{Z}_{j}}}  {\text{min}}~h(\boldsymbol{v}_{j}, \boldsymbol{v}_{j}^{t-1}), \forall{j \in \mathcal{J}},\\
      \boldsymbol{v}_{k}^{t}=\boldsymbol{v}_{k}^{t-1}, \forall{k \notin \mathcal{J}}. 
    \end{cases}       
\end{equation}

BCD algorithm can be considered as a special case of the BSUM algorithm. BSUM operation is based on successive updating blocks of primary variables with the aim of maximizing the upper-bound function of the original objective function. We can use BSUM for solving linear/ non-linear coupling constraints based on convex optimization problems that are separable. In detail, an iterative update of every variables block is performed for minimizing the upper-bound proximal function prior to converging for a joint stationary solution and piece-wise minimum. For a stationary solution to be coordinate-wise minimum, the block of variables should reach a minimum point $\boldsymbol{v}^{*}=\boldsymbol{v}_{j}^{t+1}$ \cite{tun2020joint}. \par    

\textit{Remark 1 (Convergence of BSUM Algorithm):  The BSUM algorithm takes a maximum of $\mathcal{O}(\log (1/\epsilon))$ for converging to $\epsilon$-optimal solution in a sub-linear fashion.}\par

\begin{algorithm}[t!]
	\caption{\strut BSUM Algorithm }
	\label{alg:profit}
	\begin{algorithmic}[1]
		\State{\textbf{Initialization:} Set $k=0$, $\epsilon_1>0$, and find initial feasible solutions $(\textbf{A}^{(0)}, \textbf{X}^{(0)}, \textbf{P}^{(0)})$;}
		
		\Repeat
		
		\State{Choose index set $\mathcal{I}^k$};
		\State{Let $\textbf{A}_i^{(k+1)} \in \min_{\textbf{A}_i \in \mathcal{A} } 
			\mathcal{C}_i\big(\textbf{A}_i; \textbf{A}^{(k)}, \textbf{X}^{(k)},  \textbf{P}^{(k)} \big)$};
		\State{Set $\textbf{A}_j^{(k+1)} = \ \textbf{A}_j^k$, $\forall j \notin \mathcal{I}^k$};
		\State{Find $\textbf{X}_i^{(k+1)}$, and $\textbf{P}_i^{(k+1)}$, by solving (29), and (30)};
		\State{$k = k + 1$}; 
		\Until{ $ \parallel \frac{\mathcal{C}_i^{(k)} -   \ \mathcal{C}_i^{(k+1)}}{\mathcal{C}_i^{(k)}} \parallel \ \leq \epsilon_1 $}
		
		\State{Then, set $\big(\textbf{A}_i^{(k+1)}, \textbf{X}_i^{(k+1)}, \textbf{P}_i^{(k+1)} \big)$ as the desired solution}.
	\end{algorithmic}
	\label{Algorithm}
\end{algorithm}

\subsection{Distributed Optimization Scheme for DFL Cost Minimization} 
\label{proposed algorithm}
To employ BSUM for cost minimization of DFL, we rewrite the optimization problem \textbf{P1} as follows.

\begin{align}
          \underset{\substack{ \textbf{A} \in \mathcal{A}, \textbf{X} \in \mathcal{X}, \textbf{P} \in \mathcal{P}}}{\min} \
             \mathcal{C}( \textbf{A}, \textbf{X}, \textbf{P} )   
\end{align}
where $  \mathcal{C}( \textbf{A}, \textbf{X}, \textbf{P} ) =   C_\textrm{Global}(\boldsymbol{A},\boldsymbol{X},\boldsymbol{P})$. Furthermore, the feasible of  sets of $\textbf{A}$, $\textbf{X}$, and $\textbf{P}$ are:

\begin{align*}    
\mathcal{A} \triangleq & \{ \textbf{A}: \sum_{m\in \mathcal{M}} {a_{n, m}}  \leq 1, \forall{n \in \mathcal{N}}, \sum_{n\in \mathcal{N}} {a_{n, m}}  \leq \Delta_{m}, \forall{m \in \mathcal{M}}, \\
 & a_{n, m}\in \{0,1\},\forall{n \in \mathcal{N}, m \in \mathcal{M}}\},
\end{align*} 

\begin{align*} 
\mathcal{X} \triangleq & \{ \textbf{X}: \sum_{n\in \mathcal{N}} {x_{n,r}}  \leq 1, \forall{r \in \mathcal{R}}, \sum_{r \in \mathcal{R}} {x_{n,r}}  \leq 1, \forall{n \in \mathcal{N},
 m \in \mathcal{M}}, \\ &
\sum_{r \in \mathcal{R}}\sum_{n \in \mathcal{N}} {x_{n,r}}  \leq R, x_{n,r}\in \{0,1\} \forall{n \in \mathcal{N},  m \in \mathcal{M}}\},
\end{align*} 

\begin{align*} 
\mathcal{P} \triangleq & \{ \textbf{P}: 0 \leq p_{n} \leq P_{m}, \forall{n \in \mathcal{N}}  ,\sum_{n=1}^{N}{p_{n}} \leq P_\textrm{max}^{\textrm{trans}}\}.
\end{align*} 

For the indices set $\mathcal{I}$, $k$, $\forall i \in \mathcal{I}^k$ for every iteration. The problem in (26) is still non-convex even after transforming the binary resource allocation and association variables into continuous variables. Therefore, we cannot apply a BCD-based scheme to solve it. To tackle this issue, a proximal upper-bound function $\mathcal{C}_i$ of the objective function in (26) is introduced. Specifically, we add a quadratic penalty term for penalty parameter $\mu >0$, whose basic purpose is to maintain $h$ convex.

\begin{align*}
\mathcal{C}_i(\textbf{A}_i; \textbf{A}^k, \textbf{X}^k, \textbf{P}^k) &=  \mathcal{C}(\textbf{A}_i; \tilde{\textbf{A}},  \tilde{\textbf{X}},  \tilde{\textbf{P}}) + \frac{\mu_i}{2} \parallel ( \textbf{A}_i -  \tilde{\textbf{A}}) \parallel^2.
\vspace{-0.18cm}
\end{align*}
Similarly, the quadratic penalty can be employed for $\textbf{X}_i$, and $\textbf{P}_i$. Additionally, at every iteration $h$ with respect to $\textbf{A}_i$, $\textbf{X}_i$, and $\textbf{P}_i$ in (27) yields unique $\tilde{\textbf{A}}$, $\tilde{\textbf{X}}$, and $\tilde{\textbf{P}}$, that can be taken as solution of $(k-1)$ iteration. For $(k+1)$ iteration, the solution can be given by:
 
\begin{align*}
\textbf{A}_i^{(k+1)}  & \in \min_{\textbf{A}_i \in \mathcal{A} } 
\mathcal{C}_i\bigg(\textbf{A}_i; \textbf{A}^{(k)}, \textbf{X}^{(k)}, \textbf{P}^{(k)} \bigg),  
\\
\textbf{X}_i^{(k+1)}  & \in  \min_{\textbf{X}_i \in \mathcal{X} } 
\mathcal{C}_i\bigg(\textbf{X}_i; \textbf{X}^{(k)}, \textbf{A}^{(k+1)}, \textbf{P}^ {(k)}  \bigg),  \\ 
\textbf{P}_i^{(k+1)}  & \in  \min_{\textbf{P}_i \in \mathcal{P} } 
\mathcal{C}_i \bigg(\textbf{P}_i;\textbf{P}^k, \textbf{A}^{(k+1)}, \textbf{X}^{(k+1)} \bigg),
\end{align*}
 
To solve sub-problems in (28)-(30), we use Alg.~\ref{alg:profit}.

\begin{table}[h]
	\caption{Simulation Parameters \cite{kazmi2017mode, LTE_vehicular}}
	\label{tab:SP}
	\centering
	\begin{tabular}{ll}
		\hline
		\multicolumn{1}{c}{\textbf{Simulation Parameter}}& \multicolumn{1}{c}{\textbf{Value}}    \\ \hline
		Vehicular network area & $1000\times 1000 m^{2}$ \\
		Autonomous cars & $30$ \\
     	Cellular users & $30$\\
     	Frame Structure & Type 1 (FDD) \\
		Carrier frequency (f) & 2 GHz \\
		Cars maximum transmit power & 24 dBm \\
		Thermal noise for 1 Hz at 20. C  & -174 dBm\\
			Bandwidth of resource block (W) & 180 kHz \\
		Sub carriers/resource block & 12\\

		\hline
	\end{tabular}
\end{table}

\begin{figure}[!t]
	\centering
	\captionsetup{justification=centering}
	\includegraphics[width=7cm, height=6cm]{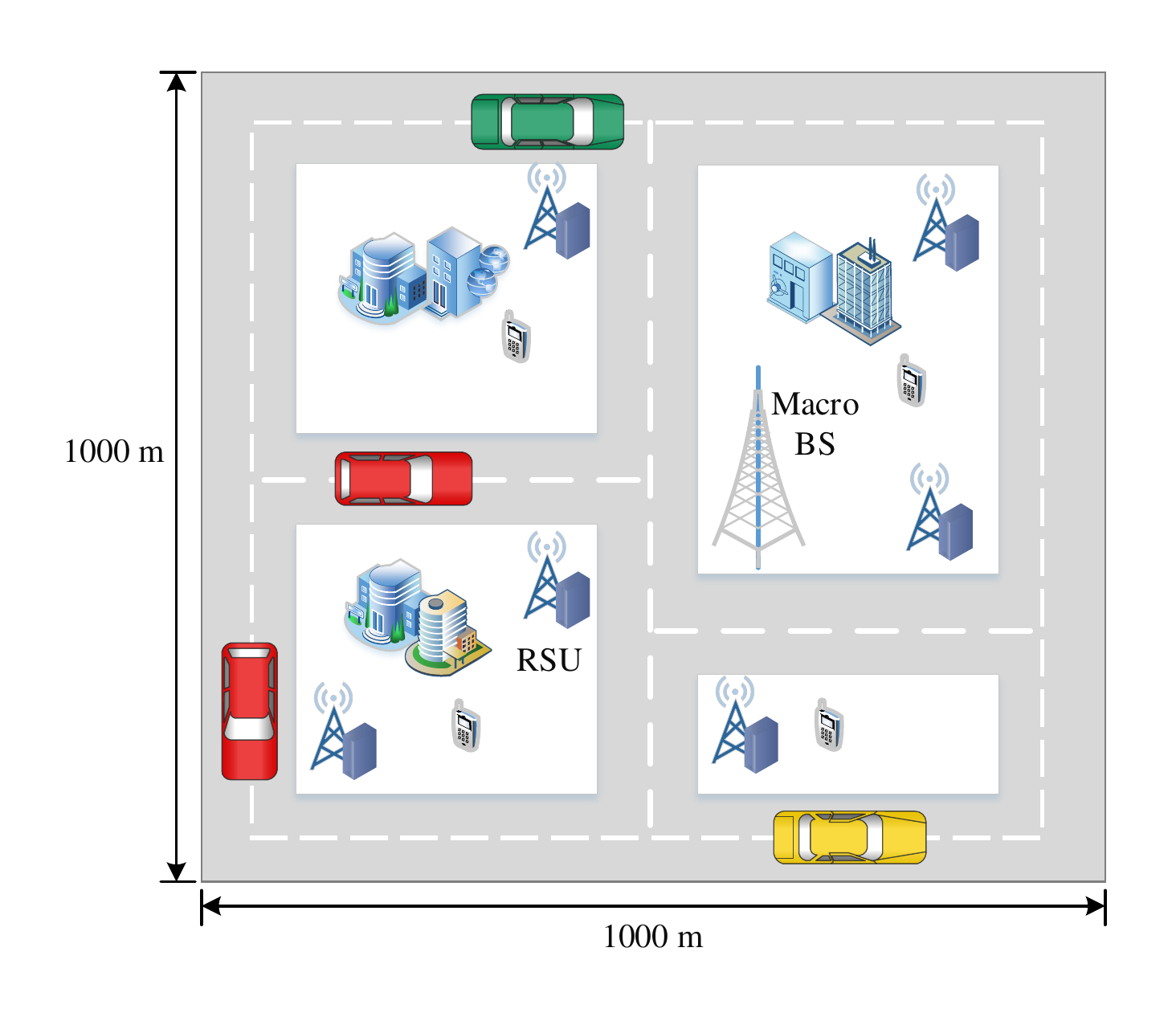}
	\caption{Vehicular network.}
	\label{fig:scenario}
\end{figure}

\section{Numerical Results}
\label{numerical results}
In this section, we present numerical results to evaluate the performance of our proposed DFL cost minimization scheme based on BSUM. We consider an area $1000 \times 1000 m^{2}$ whose layout is given in Fig.~\ref{fig:scenario}. Other simulation parameters are given in Table~\ref{tab:SP}. For performance comparison, we consider three baselines, such as baseline-A, baseline-P, and baseline-R. Baseline-A considers proposed BSUM-based association with random resource allocation and random power allocation, whereas baseline-P considers proposed BSUM-based power allocation with random resource allocation and random association. However, baseline-R considers proposed BSUM-based resource allocation with random power allocation and random association. In all results, we used an average of $30$ different runs. A sample simulation scenario consisting of cars, cellular users, and RSUs. The cellular users communicate with the central base station and we consider equal power for all cellular users. $C_\textrm{Global}$ denotes average DFL cost for all scenarios. All the cellular will occupy a single resource block. The autonomous driving cars will reuse the resource blocks by cellular users.   \par

\begin{figure*}[t!]
	\centering
		\begin{subfigure}[b]{0.3\textwidth}
		\centering
		\includegraphics[width=2.3in,height=1.8in]{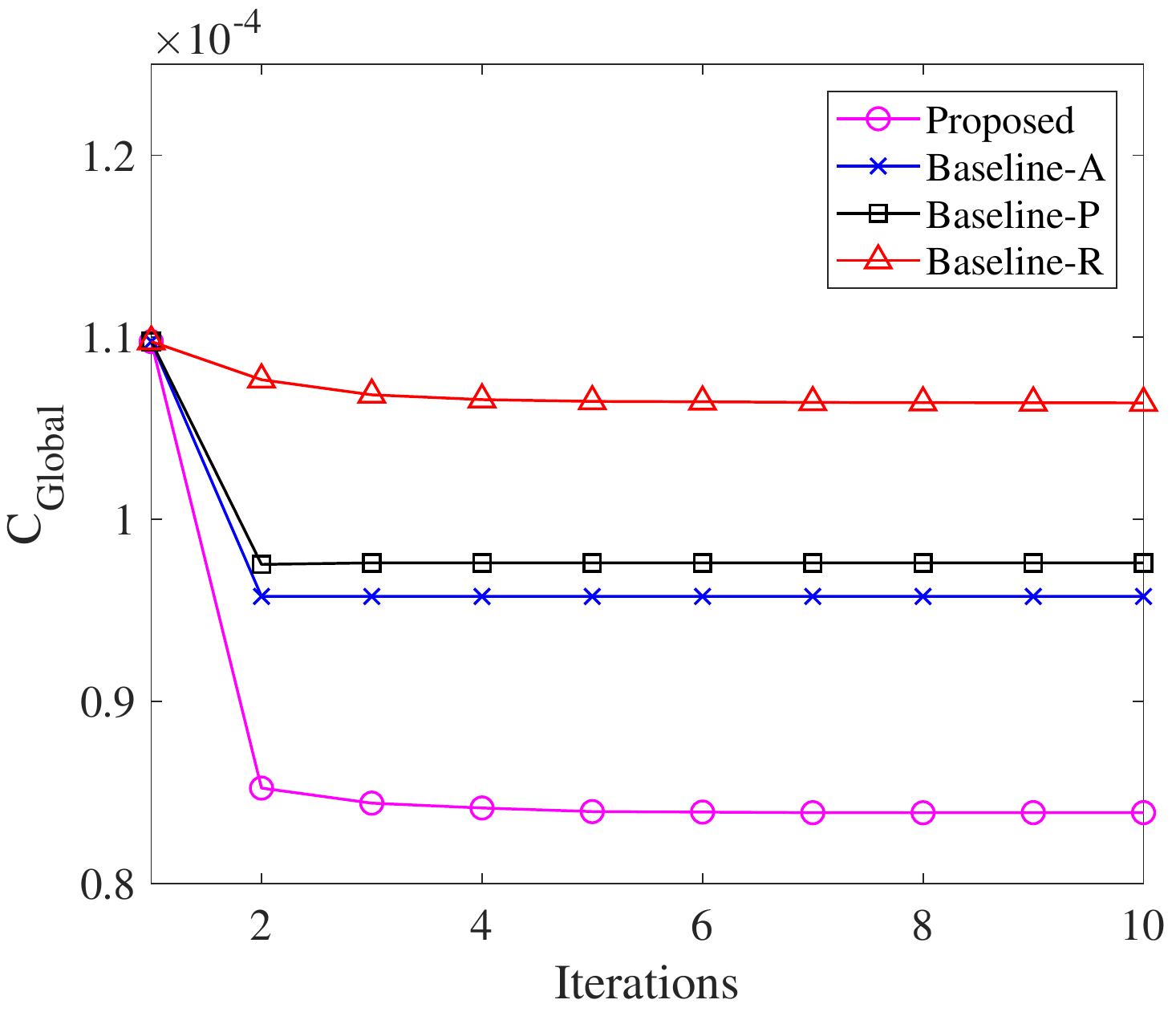}
		\caption{}	
	\end{subfigure}
	\hfill
	\begin{subfigure}[b]{0.3\textwidth}
		\centering
		\includegraphics[width=2.3in,height=1.8in]{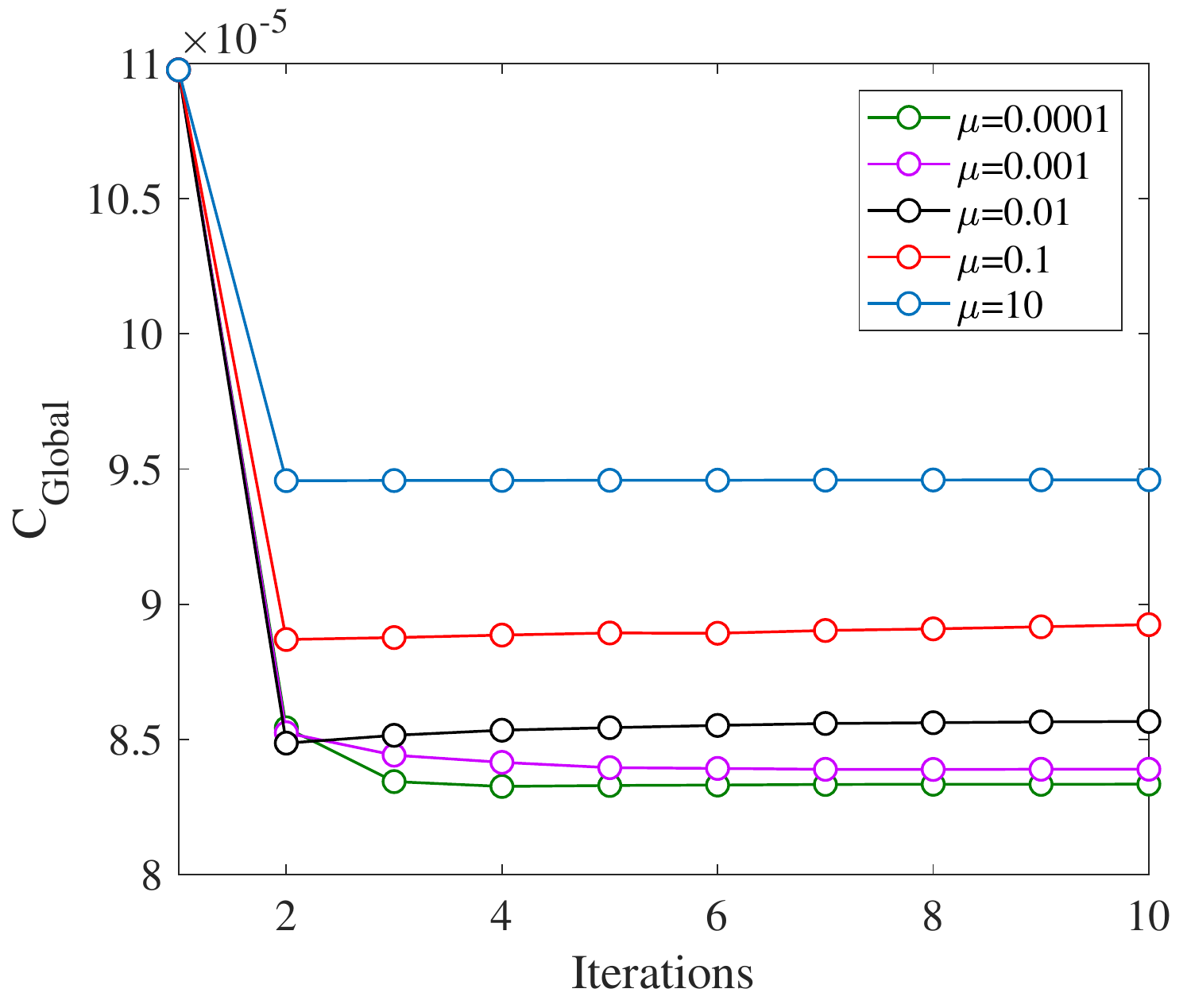}
		\caption{}	
	\end{subfigure}
	\hfill
	\begin{subfigure}[b]{0.3\textwidth}
		\centering
		\includegraphics[width=2.3in,height=1.8in]{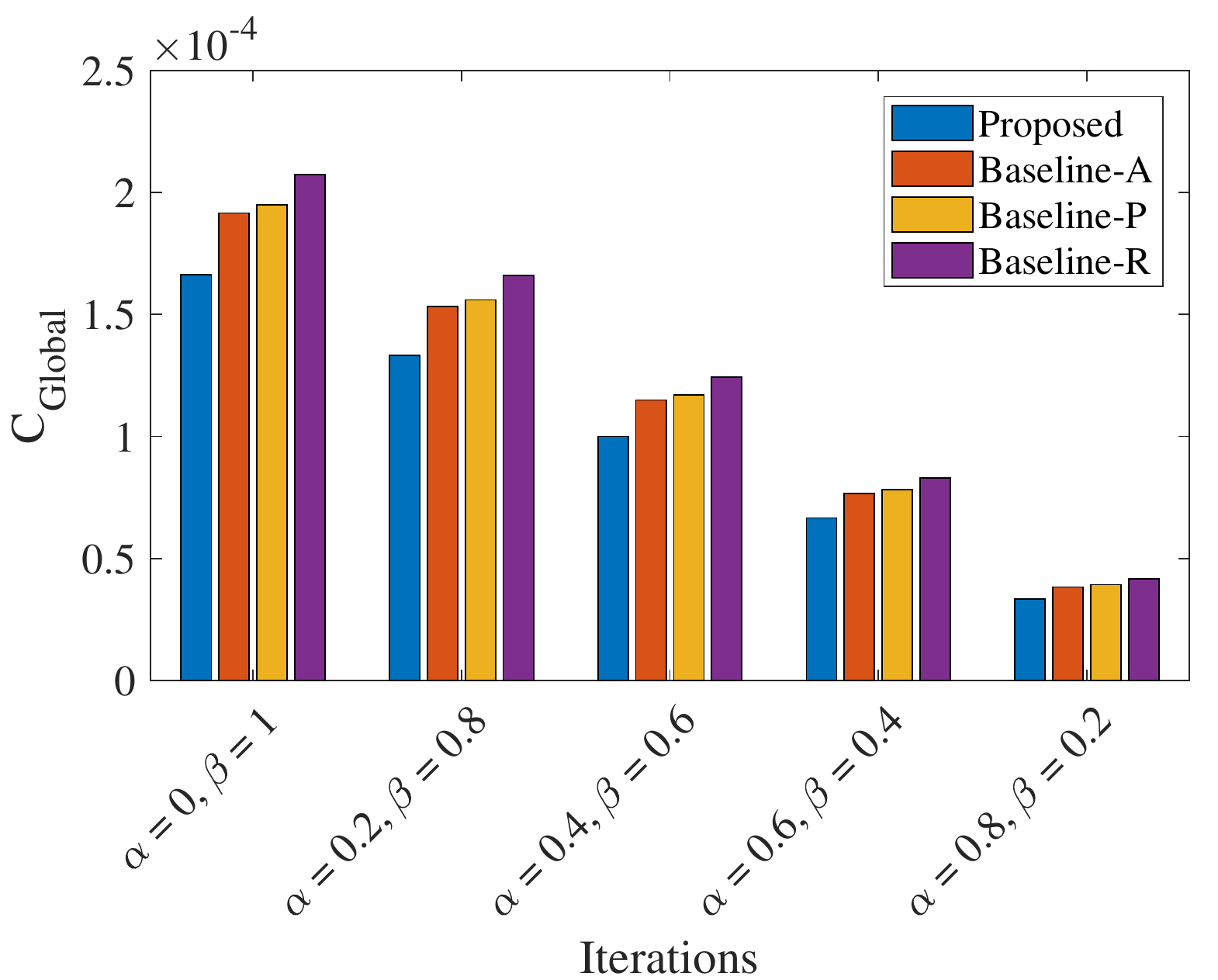}
		\caption{}
	\end{subfigure}
 
	\centering
	\caption{(a) $C_\textrm{Global}$ vs. iterations for various schemes using $\alpha=0.5$ and $\beta=0.5$, (b) $C_\textrm{Global}$ vs. iterations for different values of constants $\mu$ using $\alpha=0.5$ and $\beta=0.5$(c) $C_\textrm{Global}$ for different values of $\alpha$ and $\beta$ using $\mu=0.0001$.}
	\label{fig:simlation_3}
\end{figure*}

\begin{figure*}[t!]
	\centering
		\begin{subfigure}[b]{0.3\textwidth}
		\centering
		\includegraphics[width=2.3in,height=1.8in]{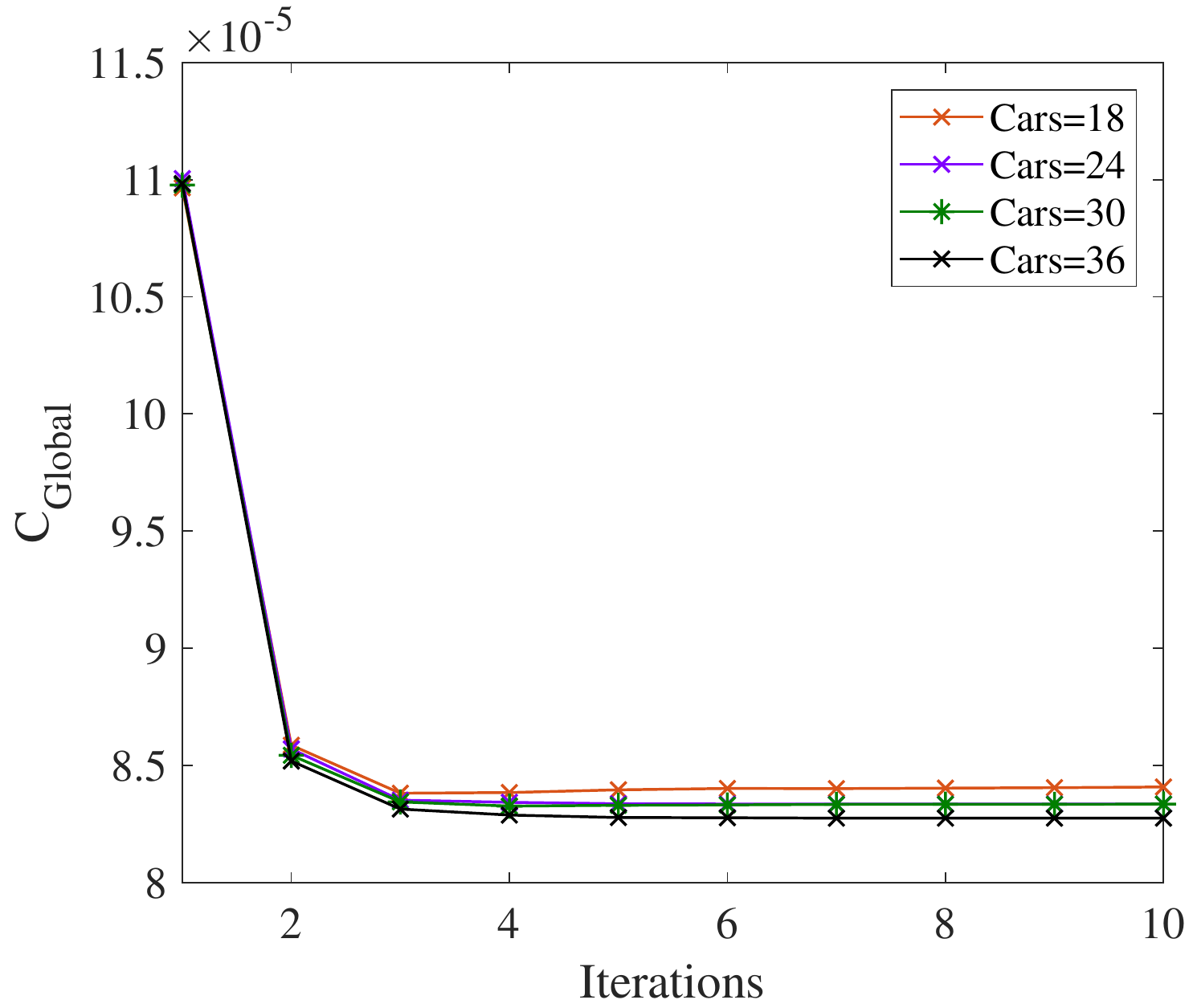}
		\caption{}	
	\end{subfigure}
	\hfill
	\begin{subfigure}[b]{0.3\textwidth}
		\centering
		\includegraphics[width=2.3in,height=1.8in]{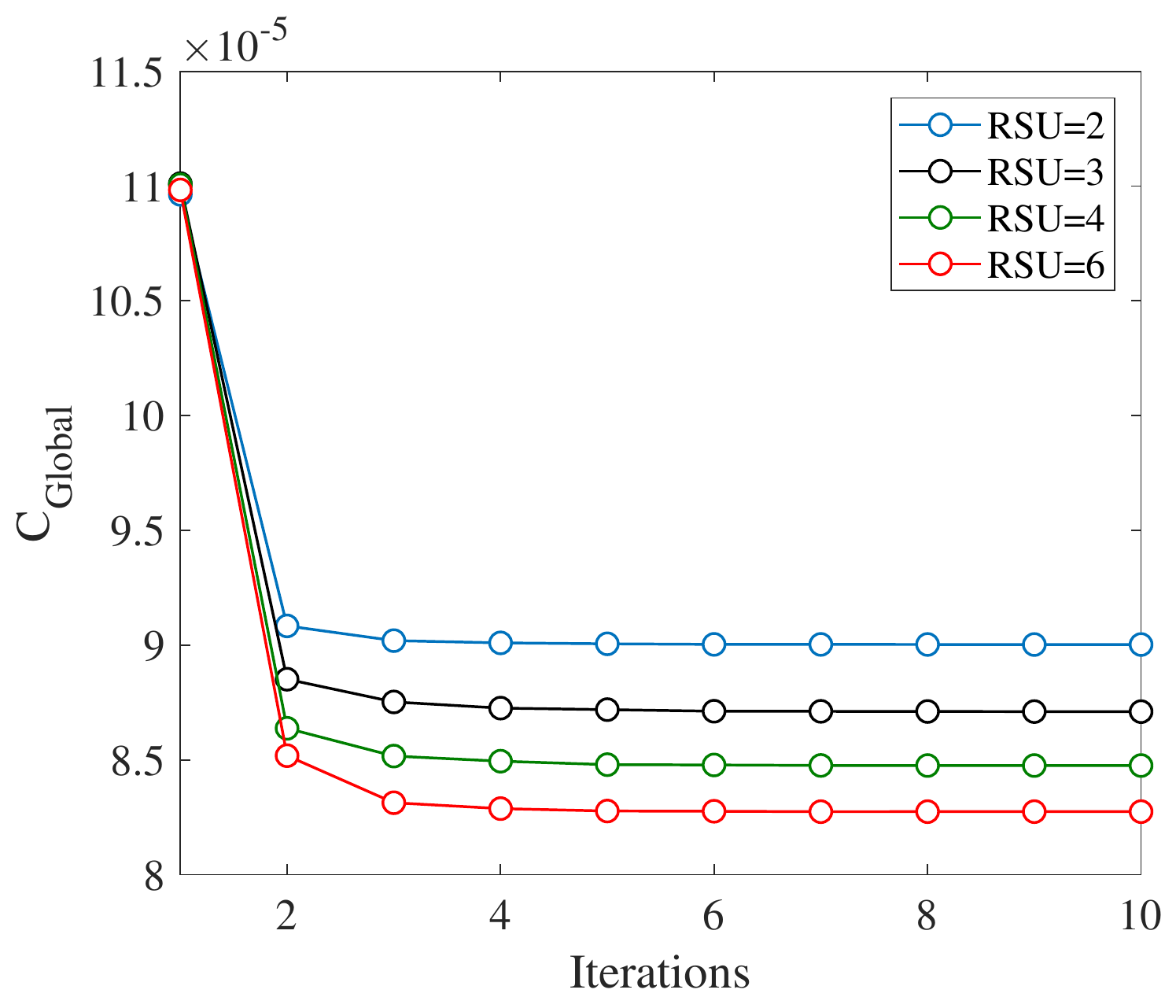}
		\caption{}	
	\end{subfigure}
	\hfill
	\begin{subfigure}[b]{0.3\textwidth}
		\centering
		\includegraphics[width=2.3in,height=1.8in]{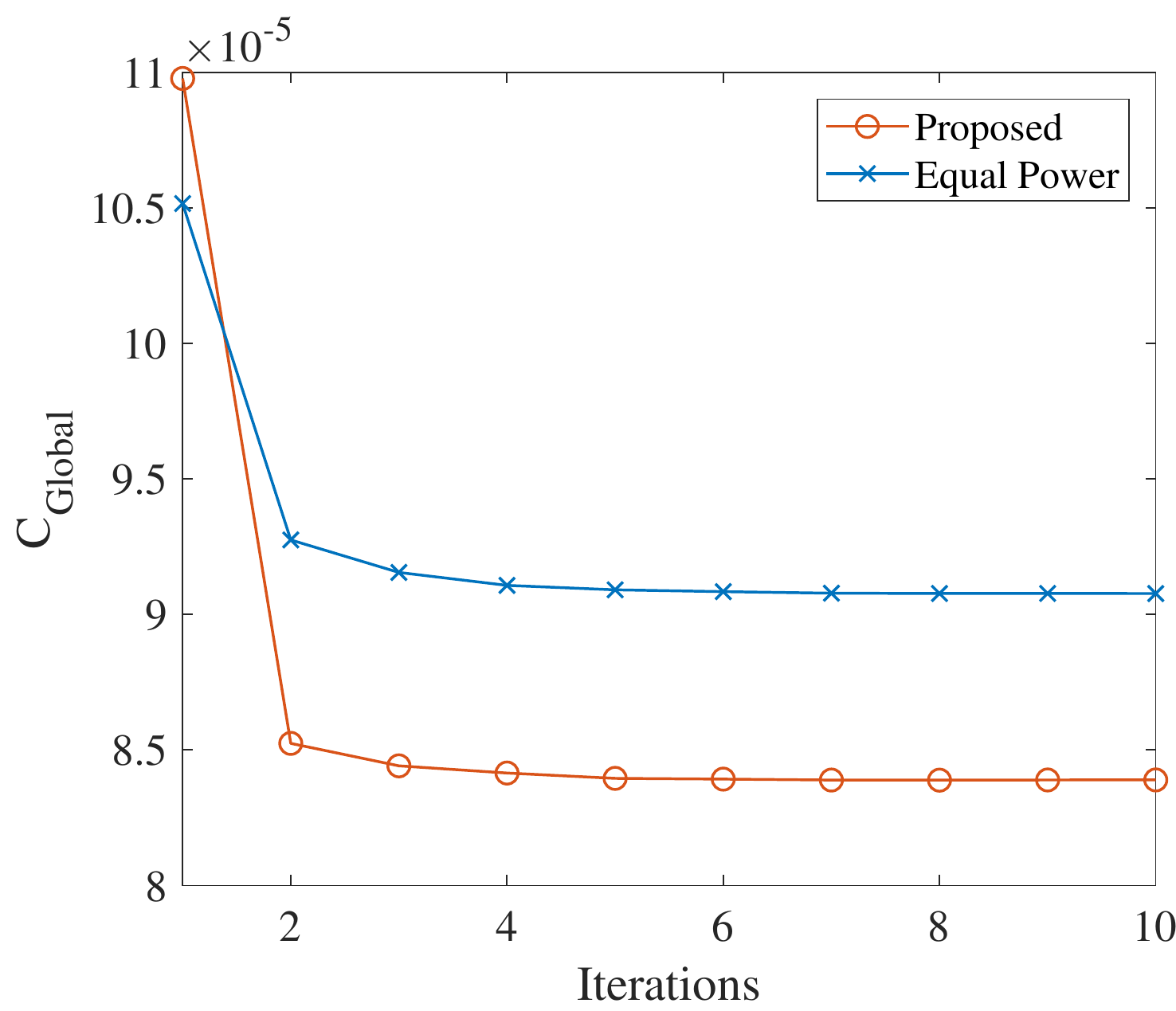}
		\caption{}
	\end{subfigure}
 
	\centering
	\caption{(a)  $C_\textrm{Global}$ vs. iterations for different cars and $6$ RSUs using $\alpha=0.5$, $\beta=0.5$, and $\mu=0.0001$, (b)  $C_\textrm{Global}$ vs. iterations for different RSUs and $36$ cars using $\alpha=0.5$, $\beta=0.5$, and $\mu=0.0001$, (c) $C_\textrm{Global}$ vs. iterations for proposed scheme and proposed scheme with equal power using $\alpha=0.5$, $\beta=0.5$, and $\mu=0.0001$.}
	\label{fig:simlation_4}
\end{figure*}
 
Fig.~\ref{fig:simlation_3}a represents $C_\textrm{Global}$ vs. iterations for various schemes, such as proposed, baseline-A, baseline-P, and baseline-R. The proposed scheme outperformed all other baseline schemes. Additionally, baseline-A has better performance than the baseline-P and baseline-R. This shows the higher impact of the association on $C_\textrm{Global}$ than power allocation and resource allocation. Baseline-R has the highest cost $C_\textrm{Global}$ which shows that resource allocation in our our scenario has the lowest impact on the $C_\textrm{Global}$. On the other hand, it is clear from Fig.~\ref{fig:simlation_3}a that all schemes converge fast (i.e., around 5 iterations). Next, we analyze the performance of $C_\textrm{Global}$ vs. iterations for different values of the constant $\mu$ in Fig.~\ref{fig:simlation_3}b. For $\mu=0.0001$, $C_\textrm{Global}$ has the lowest values which shows its preferable for use compared to other higher values of $\mu$. Fig.~\ref{fig:simlation_3}c shows the performance of various schemes for different values of scaling constants such as $\alpha$ and $\beta$ for $24$ autonomous cars and $6$ RSUs. For $\alpha=0$ and $\beta=1$, $C_\textrm{Global}$ reflects the impact of transmission latency only, whereas for $\alpha=1$ and $\beta=0$, $C_\textrm{Global}$ only shows the effect of PER on the FL learning model performance. For all values of $\alpha$ and $\beta$, it is clear from Fig.~\ref{fig:simlation_3}c that the proposed scheme outperformed all other baselines. \par   
Fig.~\ref{fig:simlation_4}a represents $C_\textrm{Global}$ vs. iterations for different number of cars using $\alpha=0.5$ and $\beta=0.5$. Generally increasing the number of cars results in slightly minimizing the cost. The reason for this is due to increasing the number of cars for a fixed number of RSUs might cause the cars to connect to nearby RSUs. Connecting to nearby RSU will result in a increasing the overall throughput, and thus lowering the value of $C_\textrm{Global}$. Fig.~\ref{fig:simlation_4}b shows $C_\textrm{Global}$ vs. iterations for different number of RSUs using fixed number of $36$ cars and $\alpha=0.5$ and $\beta=0.5$. Increasing the number of RSUs will decrease the cost of $C_\textrm{Global}$. The reason for this is due to the fact that the car will be able to connect to the nearby RSU when we increase the number of RSUs. Finally, we compare the performance of the proposed scheme (i.e., association, resource allocation, and power allocation) with the proposed scheme using fixed power in Fig.~\ref{fig:simlation_4}c. From Fig.~\ref{fig:simlation_4}c, we can see that assigning equal power to all cars will not improve the overall performance. Therefore, we must carefully assign power to cars based on the optimization algorithm (e.g., BSUM).            

\section{Conclusions}
\label{conclusion}
In this paper, we have proposed a novel DFL framework for $6$G-enabled autonomous driving cars. We have formulated a cost function that takes into accounts for transmission latency and effect of PER on FL performance. Additionally, we have formulated a MINLP optimization problem to minimize DFL cost via joint resource allocation, association, and power allocation. Due to NP-hard and non-convex nature of the formulated problem, we have proposed BSUM-based solution because of its powerful nature in solving non-convex, complex problems. In addition, we have provided extensive numerical results to validate our proposal. We concluded that DFL can be employed for various intelligent transportation applications by providing FL in a robust, resource efficient, and privacy-aware manner. Furthermore, we concluded that DFL can serve as a practical application of FL for other $6$G applications.

\bibliographystyle{IEEEtran}
\bibliography{Database}

\begin{thebibliography}{10}
\providecommand{\url}[1]{#1}
\csname url@samestyle\endcsname
\providecommand{\newblock}{\relax}
\providecommand{\bibinfo}[2]{#2}
\providecommand{\BIBentrySTDinterwordspacing}{\spaceskip=0pt\relax}
\providecommand{\BIBentryALTinterwordstretchfactor}{4}
\providecommand{\BIBentryALTinterwordspacing}{\spaceskip=\fontdimen2\font plus
\BIBentryALTinterwordstretchfactor\fontdimen3\font minus
  \fontdimen4\font\relax}
\providecommand{\BIBforeignlanguage}[2]{{%
\expandafter\ifx\csname l@#1\endcsname\relax
\typeout{** WARNING: IEEEtran.bst: No hyphenation pattern has been}%
\typeout{** loaded for the language `#1'. Using the pattern for}%
\typeout{** the default language instead.}%
\else
\language=\csname l@#1\endcsname
\fi
#2}}
\providecommand{\BIBdecl}{\relax}
\BIBdecl

\bibitem{9163104}
N.~{Kato}, B.~{Mao}, F.~{Tang}, Y.~{Kawamoto}, and J.~{Liu}, ``Ten challenges
  in advancing machine learning technologies toward 6{G},'' \emph{IEEE Wireless
  Communications}, vol.~27, no.~3, pp. 96--103, June 2020.

\bibitem{akyildiz20206g}
I.~F. {Akyildiz}, A.~{Kak}, and S.~{Nie}, ``6{G} and beyond: The future of
  wireless communications systems,'' \emph{IEEE Access}, vol.~8, pp.
  133\,995--134\,030, July 2020.

\bibitem{yang20196g}
P.~Yang, Y.~Xiao, M.~Xiao, and S.~Li, ``6{G} wireless communications: Vision
  and potential techniques,'' \emph{IEEE Network}, vol.~33, no.~4, pp. 70--75,
  July 2019.

\bibitem{khan2021digital}
L.~U. Khan, W.~Saad, D.~Niyato, Z.~Han, and C.~S. Hong, ``Digital-twin-enabled
  6g: Vision, architectural trends, and future directions,'' \emph{arXiv
  preprint arXiv:2102.12169}, 2021.

\bibitem{saad2019vision}
W.~Saad, M.~Bennis, and M.~Chen, ``A vision of 6g wireless systems:
  Applications, trends, technologies, and open research problems,'' \emph{IEEE
  network}, vol.~34, no.~3, pp. 134--142, 2019.

\bibitem{grigorescu2019survey}
S.~Grigorescu, B.~Trasnea, T.~Cocias, and G.~Macesanu, ``A survey of deep
  learning techniques for autonomous driving,'' \emph{Journal of Field
  Robotics}, 2019.

\bibitem{ndikumana2020deep}
A.~Ndikumana, N.~H. Tran, K.~T. Kim, C.~S. Hong \emph{et~al.}, ``Deep learning
  based caching for self-driving cars in multi-access edge computing,''
  \emph{Early access, IEEE Transactions on Intelligent Transportation Systems},
  2020.

\bibitem{mcmahan2016communication}
H.~B. McMahan, E.~Moore, D.~Ramage, S.~Hampson \emph{et~al.},
  ``Communication-efficient learning of deep networks from decentralized
  data,'' \emph{arXiv preprint arXiv:1602.05629}, 2016.

\bibitem{khan2019federated}
L.~U. {Khan}, S.~R. {Pandey}, N.~H. {Tran}, W.~{Saad}, Z.~{Han}, M.~N.~H.
  {Nguyen}, and C.~S. {Hong}, ``Federated learning for edge networks: Resource
  optimization and incentive mechanism,'' \emph{IEEE Communications Magazine},
  vol.~58, no.~10, pp. 88--93, October 2020.

\bibitem{FL_industry_1}
``Federated learning, a step closer towards confidential ai,''
  \url{https://medium.com/frstvc/tagged/thoughts}, [Online; accessed Jan. 24,
  2020].

\bibitem{khan2019edge}
L.~U. {Khan}, I.~{Yaqoob}, N.~H. {Tran}, S.~M.~A. {Kazmi}, T.~N. {Dang}, and
  C.~S. {Hong}, ``Edge-computing-enabled smart cities: A comprehensive
  survey,'' \emph{IEEE Internet of Things Journal}, vol.~7, no.~10, pp.
  10\,200--10\,232, October 2020.

\bibitem{khan2020federated}
L.~U. Khan, W.~Saad, Z.~Han, E.~Hossain, and C.~S. Hong, ``Federated learning
  for internet of things: Recent advances, taxonomy, and open challenges,''
  \emph{arXiv preprint arXiv:2009.13012}, 2020.

\bibitem{lim2019federated}
W.~Y.~B. Lim, N.~C. Luong, D.~T. Hoang, Y.~Jiao, Y.-C. Liang, Q.~Yang,
  D.~Niyato, and C.~Miao, ``Federated learning in mobile edge networks: A
  comprehensive survey,'' \emph{arXiv preprint arXiv:1909.11875}, 2019.

\bibitem{kim2019blockchained}
H.~{Kim}, J.~{Park}, M.~{Bennis}, and S.~{Kim}, ``Blockchained on-device
  federated learning,'' \emph{IEEE Communications Letters}, vol.~24, no.~6, pp.
  1279--1283, June 2020.

\bibitem{abad2019hierarchical}
M.~S.~H. {Abad}, E.~{Ozfatura}, D.~{GUndUz}, and O.~{Ercetin}, ``Hierarchical
  federated learning across heterogeneous cellular networks,'' pp. 8866--8870,
  May 2020.

\bibitem{seif2020wireless}
M.~Seif, R.~Tandon, and M.~Li, ``Wireless federated learning with local
  differential privacy,'' \emph{arXiv preprint arXiv:2002.05151}, 2020.

\bibitem{khan2020dispersed}
L.~U. Khan, W.~Saad, Z.~Han, and C.~S. Hong, ``Dispersed federated learning:
  Vision, taxonomy, and future directions,'' \emph{arXiv preprint
  arXiv:2008.05189}, 2020.

\bibitem{dinh2019federated}
N.~H. Tran, W.~Bao, A.~Zomaya, and C.~S. Hong, ``Federated learning over
  wireless networks: Optimization model design and analysis,'' pp. 1387--1395,
  May 2019.

\bibitem{chen2019joint}
M.~Chen, Z.~Yang, W.~Saad, C.~Yin, H.~V. Poor, and S.~Cui, ``A joint learning
  and communications framework for federated learning over wireless networks,''
  \emph{arXiv preprint arXiv:1909.07972}, 2019.

\bibitem{wang2019adaptive}
S.~Wang, T.~Tuor, T.~Salonidis, K.~K. Leung, C.~Makaya, T.~He, and K.~Chan,
  ``Adaptive federated learning in resource constrained edge computing
  systems,'' \emph{IEEE Journal on Selected Areas in Communications}, vol.~37,
  no.~6, pp. 1205--1221, March 2019.

\bibitem{ndikumana2019joint}
A.~Ndikumana, N.~H. Tran, T.~M. Ho, Z.~Han, W.~Saad, D.~Niyato, and C.~S. Hong,
  ``Joint communication, computation, caching, and control in big data
  multi-access edge computing,'' \emph{IEEE Transactions on Mobile Computing},
  vol.~19, no.~6, pp. 1359--1374, June 2019.

\bibitem{tun2020joint}
Y.~K. Tun, A.~Ndikumana, S.~R. Pandey, Z.~Han, and C.~S. Hong, ``Joint radio
  resource allocation and content caching in heterogeneous virtualized wireless
  networks,'' \emph{IEEE Access}, vol.~8, pp. 36\,764--36\,775, February 2020.

\bibitem{hong2015unified}
M.~Hong, M.~Razaviyayn, Z.-Q. Luo, and J.-S. Pang, ``A unified algorithmic
  framework for block-structured optimization involving big data: With
  applications in machine learning and signal processing,'' \emph{IEEE Signal
  Processing Magazine}, vol.~33, no.~1, pp. 57--77, January 2015.

\bibitem{Client-Edge}
L.~{Liu}, J.~{Zhang}, S.~H. {Song}, and K.~B. {Letaief}, ``Client-edge-cloud
  hierarchical federated learning,'' in \emph{IEEE International Conference on
  Communications}, July 2020, pp. 1--6.

\bibitem{xi2011general}
Y.~Xi, A.~Burr, J.~Wei, and D.~Grace, ``A general upper bound to evaluate
  packet error rate over quasi-static fading channels,'' \emph{IEEE
  Transactions on Wireless Communications}, vol.~10, no.~5, pp. 1373--1377,
  January 2011.

\bibitem{friedlander2012hybrid}
M.~P. Friedlander and M.~Schmidt, ``Hybrid deterministic-stochastic methods for
  data fitting,'' \emph{SIAM Journal on Scientific Computing}, vol.~34, no.~3,
  pp. A1380--A1405, May 2012.

\bibitem{tun2020energy}
Y.~K. Tun, Y.~M. Park, N.~H. Tran, W.~Saad, S.~R. Pandey, and C.~S. Hong,
  ``Energy-efficient resource management in uav-assisted mobile edge
  computing,'' \emph{Early access, IEEE Communications Letters}, 2020.

\bibitem{han2017signal}
Z.~Han, M.~Hong, and D.~Wang, \emph{Signal processing and networking for big
  data applications}.\hskip 1em plus 0.5em minus 0.4em\relax Cambridge
  University Press, 2017.

\bibitem{kazmi2017mode}
S.~A. Kazmi, N.~H. Tran, W.~Saad, Z.~Han, T.~M. Ho, T.~Z. Oo, and C.~S. Hong,
  ``Mode selection and resource allocation in device-to-device communications:
  A matching game approach,'' \emph{IEEE Transactions on Mobile Computing},
  vol.~16, no.~11, pp. 3126--3141, March 2017.

\bibitem{LTE_vehicular}
G.~{Araniti}, C.~{Campolo}, M.~{Condoluci}, A.~{Iera}, and A.~{Molinaro}, ``Lte
  for vehicular networking: a survey,'' \emph{IEEE Communications Magazine},
  vol.~51, no.~5, pp. 148--157, May 2013.

\end{thebibliography}

\begin{IEEEbiography}[{\includegraphics[width=1in,height=1.25in,clip,keepaspectratio]{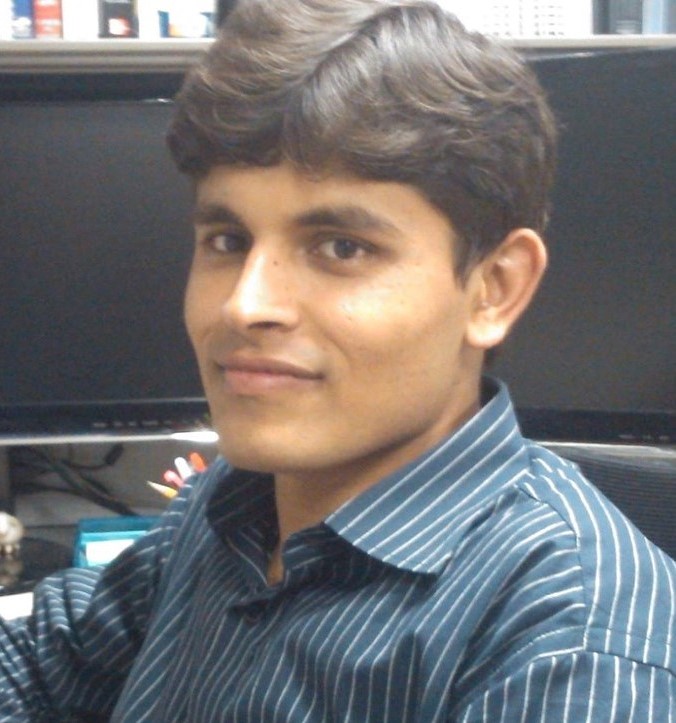}}]{Latif U. Khan} is currently pursuing his Ph.D. degree in Computer Engineering at Kyung Hee University (KHU), South Korea. He is working as a leading researcher in the intelligent Networking Laboratory under a project jointly funded by the prestigious Brain Korea 21st Century Plus and Ministry of Science and ICT, South Korea. He received his MS (Electrical Engineering) degree with distinction from University of Engineering and Technology (UET), Peshawar, Pakistan in 2017. Prior to joining the KHU, he has served as a faculty member and research associate in the UET, Peshawar, Pakistan. He is the author/co-author of two conference best paper awards. He is also author of the book "Network Slicing for $5$G and Beyond Networks". His research interests include analytical techniques of optimization and game theory to edge computing, end-to-end network slicing, and federated learning for wireless networks. \end{IEEEbiography}

\begin{IEEEbiography}[{\includegraphics[width=1in,height=1.3in,clip,keepaspectratio]{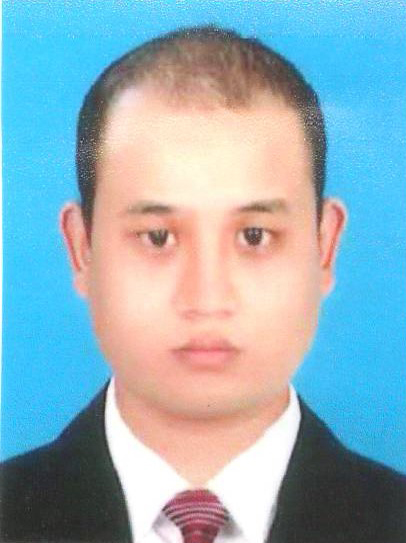}}]{Yan Kyaw Tun} received the B.E. degree in marine electrical systems and electronics engineering from Myanmar Maritime University, Thanlyin, Myanmar, in 2014. He is currently pursuing the Ph.D. degree in computer science and engineering with Kyung Hee University, South Korea, for which he received a scholarship, in 2015. His research interests include network economics, game theory, network optimization, wireless communication, wireless network virtualization, mobile edge computing, and wireless resource slicing for 5G.
\end{IEEEbiography} 

\begin{IEEEbiography}[{\includegraphics[width=1in,height=1.25in,clip,keepaspectratio]{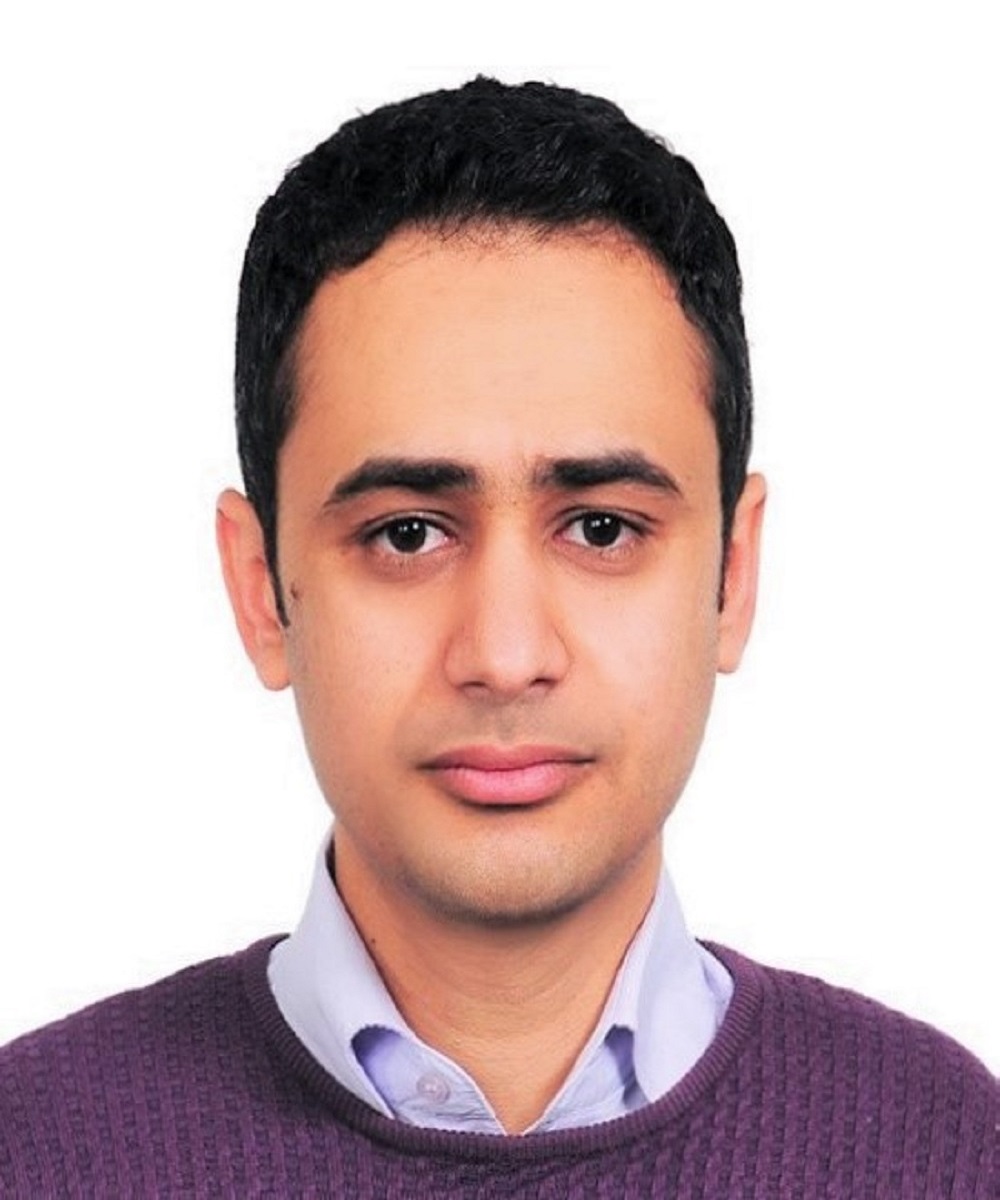}}]{Madyan Alsenwi} received the B.E. and M.Sc. degrees in electronics and communications engineering from Cairo University, Egypt, in 2011 and 2016, respectively. He is working as a Leading Researcher at the Intelligent Networking Laboratory, Department of Computer Science and Engineering, Kyung Hee University under a project jointly funded by the prestigious Brain Korea 21st Century Plus and Ministry of Science and ICT, South Korea. Prior to this, he worked as a Research Assistant under several research projects funded by the Information Technology Industry Development Agency (ITIDA), Egypt. His research interests include resource management in 5G networks and beyond, ultra-reliable low latency communications (URLLC), UAV-assisted wireless networks, and learning over wireless networks.
\end{IEEEbiography}

\begin{IEEEbiography}[{\includegraphics[width=1in,height=1.25in,clip,keepaspectratio]{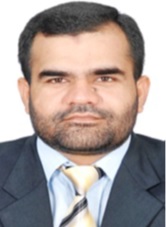}}]
{Muhammad Imran} is an associate professor at King Saud University. His research interest includes MANET, WSNs, WBANs, M2M/IoT, SDN, Security and privacy. He has published a number of research papers in refereed international conferences and journals. He served as a Co-Editor in Chief for EAI Transactions and Associate/Guest editor for IEEE (Access, Communications, Wireless Communications Magazine), Future Generation Computer Systems, Computer Networks, Sensors, IJDSN, JIT, WCMC, AHSWN, IET WSS, IJAACS and IJITEE.
\end{IEEEbiography} \par

\begin{IEEEbiography}[{\includegraphics[width=1in,height=1.25in,clip,keepaspectratio]{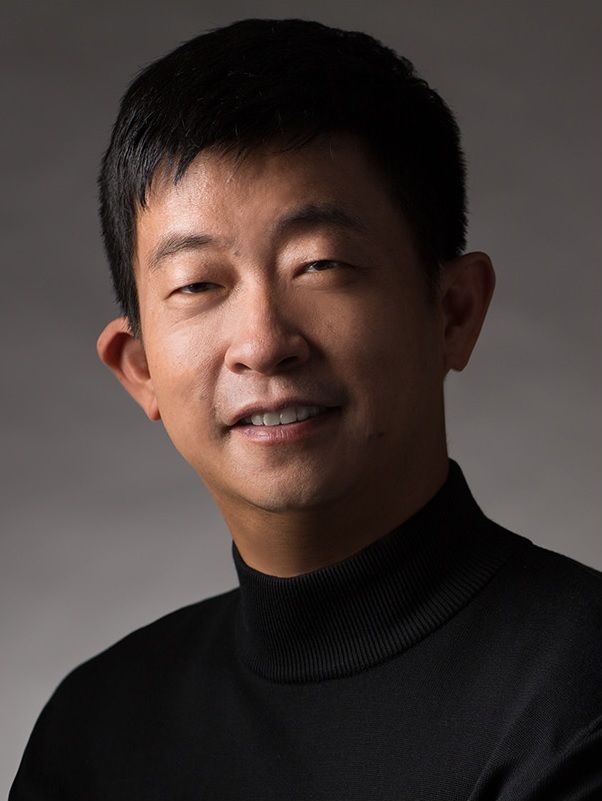}}]{Zhu Han}(S’01, M’04, SM’09, F’14) received the B.S. degree in electronic engineering from Tsinghua University, in 1997, and the M.S. and Ph.D. degrees in electrical and computer engineering from the University of Maryland, College Park, in 1999 and 2003, respectively. From 2000 to 2002, he was an R\&D Engineer of JDSU, Germantown, Maryland. From 2003 to 2006, he was a Research Associate at the University of Maryland. From 2006 to 2008, he was an assistant professor at Boise State University, Idaho. Currently, he is a John and Rebecca Moores Professor in the Electrical and Computer Engineering Department as well as in the Computer Science Department at the University of Houston, Texas. He is also a Chair professor in National Chiao Tung University, ROC. His research interests include wireless resource allocation and management, wireless communications and networking, game theory, big data analysis, security, and smart grid.  Dr. Han received an NSF Career Award in 2010, the Fred W. Ellersick Prize of the IEEE Communication Society in 2011, the EURASIP Best Paper Award for the Journal on Advances in Signal Processing in 2015, IEEE Leonard G. Abraham Prize in the field of Communications Systems (best paper award in IEEE JSAC) in 2016, and several best paper awards in IEEE conferences. Dr. Han was an IEEE Communications Society Distinguished Lecturer from 2015-2018, and is AAAS fellow since 2019 and ACM distinguished Member since 2019. Dr. Han is 1\% highly cited researcher since 2017 according to Web of Science.\end{IEEEbiography}


\begin{IEEEbiography}[{\includegraphics[width=1in,height=1.25in,clip,keepaspectratio]{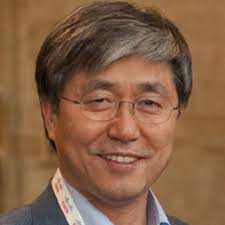}}]{Choong Seon Hong} (S’95-M’97-SM’11)  received the B.S. and M.S. degrees in electronic engineering from Kyung Hee University, Seoul, South Korea, in 1983 and 1985, respectively, and the Ph.D. degree from Keio University, Tokyo, Japan, in 1997. In 1988, he joined KT, Gyeonggi-do, South Korea, where he was involved in broadband networks as a member of the Technical Staff. Since 1993, he has been with Keio University. He was with the Telecommunications Network Laboratory, KT, as a Senior Member of Technical Staff and as the Director of the Networking Research Team until 1999. Since 1999, he has been a Professor with the Department of Computer Science and Engineering, Kyung Hee University. His research interests include future Internet, intelligent edge computing, network management, and network security.  Dr. Hong is a member of the Association for Computing Machinery (ACM), the Institute of Electronics, Information and Communication Engineers (IEICE), the Information Processing Society of Japan (IPSJ), the Korean Institute of Information Scientists and Engineers (KIISE), the Korean Institute of Communications and Information Sciences (KICS), the Korean Information Processing Society (KIPS), and the Open Standards and ICT Association (OSIA). He has served as the General Chair, the TPC Chair/Member, or an Organizing Committee Member of international conferences, such as the Network Operations and Management Symposium (NOMS), International Symposium on Integrated Network Management (IM), Asia-Pacific Network Operations and Management Symposium (APNOMS), End-to-End Monitoring Techniques and Services (E2EMON), IEEE Consumer Communications and Networking Conference (CCNC), Assurance in Distributed Systems and Networks (ADSN), International Conference on Parallel Processing (ICPP), Data Integration and Mining (DIM), World Conference on Information Security Applications (WISA), Broadband Convergence Network (BcN), Telecommunication Information Networking Architecture (TINA), International Symposium on Applications and the Internet (SAINT), and International Conference on Information Networking (ICOIN). He was an Associate Editor of the IEEE TRANSACTIONS ON NETWORK AND SERVICE MANAGEMENT and the IEEE JOURNAL OF COMMUNICATIONS AND NETWORKS. He currently serves as an Associate Editor for the International Journal of Network Management.\end{IEEEbiography}

\end{document}